\pdfoutput=1
\documentclass[11pt,twoside,a4paper,cmspaper,final,collab]{cms-tdr}
\def\svnVersion{e9ff110}\def\svnDate{2021/08/20}\def\cmsCernNoTag{CERN-EP-2021-052}\def\cmsCernDate{\today}\def\cmsMessage{Published in Physical Review D as \href{http://dx.doi.org/10.1103/PhysRevD.104.052011}{\doi{10.1103/PhysRevD.104.052011}.}}
\begin{document}\cmsNoteHeader{EXO-19-013}

\cmsNoteHeader{EXO-19-013}

\newcommand{\fb}{\unit{fb}}
\newcommand{\intlumiTotal}{140\fbinv}
\newcommand{\intlumiFirst}{38.5\fbinv}
\newcommand{\intlumiSecond}{101\fbinv}
\newcommand{\dxy}{\ensuremath{d_{xy}}\xspace}
\newcommand{\dbv}{\ensuremath{d_{\mathrm{BV}}}\xspace}
\newcommand{\dvv}{\ensuremath{d_{\mathrm{VV}}}\xspace}
\newcommand{\dvvc}{\ensuremath{d_{\mathrm{VV}}^{\kern 0.15em\mathrm{C}}}\xspace}
\newcommand{\dphivv}{\ensuremath{\Delta\phi_{\mathrm{VV}}}\xspace}
\newcommand{\dphijj}{\ensuremath{\Delta\phi_{\mathrm{JJ}}}\xspace}
\newcommand{\nsigmadxy}{\ensuremath{\dxy/\sigma_{d_{xy}}}\xspace}
\newcommand{\sigmaBsq}{\sigma\mathcal{B}^2}
\newlength\cmsTabSkip\setlength{\cmsTabSkip}{1ex}
\providecommand{\cmsTable}[1]{\resizebox{\textwidth}{!}{#1}}
\ifthenelse{\boolean{cms@external}}{\providecommand{\cmsLeft}{upper\xspace}}{\providecommand{\cmsLeft}{left\xspace}}
\ifthenelse{\boolean{cms@external}}{\providecommand{\cmsRight}{lower\xspace}}{\providecommand{\cmsRight}{right\xspace}}

\title{Search for long-lived particles decaying to jets with displaced vertices in proton-proton collisions at \texorpdfstring{$\sqrt{s}=13\TeV$}{sqrt(s) = 13 TeV}}

\date{\today}

\abstract{
A search is presented for long-lived particles produced in pairs in proton-proton collisions at the LHC operating at a center-of-mass energy of 13\TeV. The data were collected with the CMS detector during the period from 2015 through 2018, and correspond to a total integrated luminosity of \intlumiTotal. This search targets pairs of long-lived particles with mean proper decay lengths between 0.1 and 100\mm, each of which decays into at least two quarks that hadronize to jets, resulting in a final state with two displaced vertices. No significant excess of events with two displaced vertices is observed. In the context of $R$-parity violating supersymmetry models, the pair production of long-lived neutralinos, gluinos, and top squarks is excluded at 95\% confidence level for cross sections larger than 0.08\fb, masses between 800 and 3000\GeV, and mean proper decay lengths between 1 and 25\mm. 
}

\hypersetup{
pdfauthor={CMS Collaboration},
pdftitle={Search for long-lived particles decaying to jets with displaced vertices in proton-proton collisions at sqrt(s)= 13 TeV},
pdfsubject={CMS},
pdfkeywords={CMS, displaced vertices}}

\maketitle

\section{Introduction}
Particles with lifetimes corresponding to macroscopically long decay lengths are common in models of physics beyond the standard model (SM). Models predicting the production of long-lived particles at the CERN LHC include $R$-parity violating (RPV) supersymmetry (SUSY)~\cite{Barbier:2004ez,Yuval,Csaba,Hall:1983id}, split SUSY~\cite{ArkaniHamed:2004fb,ArkaniHamed:2004yi,ArkaniHamed:2012gw,Arvanitaki:2012ps,Gambino:2005eh,Giudice:2004tc,Hewett:2004nw}, hidden-valley models~\cite{Han:2007ae,Strassler:2006im,Strassler:2006ri}, stealth SUSY~\cite{Fan:2011yu,Fan:2012jf}, and other models giving rise to dark matter candidates~\cite{Calibbi:2018fqf,Co:2015pka,Cui:2011ab,Cui:2012jh,Cui:2014twa,Hall:2009bx,Kaplan:2009ag,Kim:2013ivd}. Searches for long-lived particles, therefore, probe a large beyond-the-SM parameter space.

The broad parameter space calls for an inclusive and model-independent search. This analysis searches for long-lived particles that are produced in pairs and decay into final states with multiple jets containing charged particles. Specifically, the analysis looks for a unique experimental signature consisting of two vertices, each formed from the intersection of multiple charged-particle trajectories and displaced from the interaction region but within the radius of the beam pipe (22\mm). 

\begin{figure}[hbtp]
\centering
\includegraphics[width=0.49\textwidth]{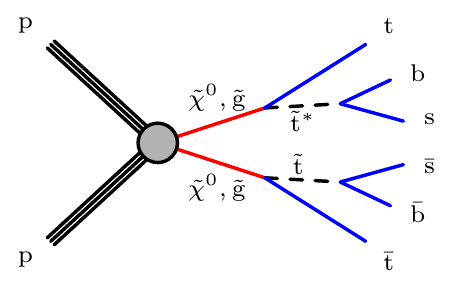}
\includegraphics[width=0.49\textwidth]{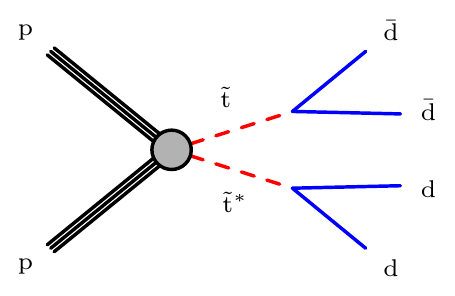}
\caption{Diagrams of the multijet signal model (\cmsLeft) showing long-lived neutralinos (\PSGcz) or gluinos (\PSg) decaying into top, bottom, and strange quarks via virtual top squarks (\PSQt), and the dijet signal model (\cmsRight) showing long-lived top and anti-top squarks decaying into two down-type quarks. In both cases, the long-lived particles are the LSPs in their respective models.}
\label{fig:diagrams}
\end{figure}

This analysis uses as benchmarks two signal models with distinct final states. The first is a minimal flavor violating model of RPV SUSY~\cite{Yuval} in which the lightest SUSY particle (LSP) is a long-lived neutralino or gluino, either of which is pair produced. The long-lived particle then decays into top, bottom, and strange quarks, as shown in Fig.~\ref{fig:diagrams} (\cmsLeft), resulting in a ``multijet'' final-state signal topology. The second benchmark model is another RPV model in which the pair-produced top squark is the long-lived LSP~\cite{Csaba}. Each squark decays into a pair of down-type quarks, resulting in a ``dijet'' final state signature, shown in Fig.~\ref{fig:diagrams} (\cmsRight).

The displaced vertices are reconstructed from charged-particle tracks using a custom vertex reconstruction algorithm.  To discriminate the signal from SM backgrounds, we use the separation between the vertex pairs in the plane transverse to the beam direction. Signal events tend to have well-separated vertex pairs, while background events, whose vertices originate from track misreconstruction and therefore tend to cluster near the beam axis, typically exhibit little vertex separation.

We target signals with lifetimes corresponding to mean proper decay lengths ($c\tau$) in the range from 0.1 to 100\mm. This search is primarily sensitive to models in which the mass of each long-lived particle exceeds approximately 600\GeV because of a trigger requiring large total jet transverse momentum. 

The previous CMS displaced vertex search~\cite{CMS-EXO-17-018} was based on data collected in 2015 and 2016. This analysis is an extension of that search, utilizing events collected in 2017 and 2018, corresponding to an integrated luminosity of $101\fbinv$. We then determine results based on the full Run-2 dataset, which spans from 2015 to 2018 and corresponds to a total integrated luminosity of \intlumiTotal. The CMS Collaboration upgraded its pixel tracking detector during the winter technical stop between the 2016 and 2017 running periods~\cite{CMS:2012sda}, providing improvements that benefit this analysis, which relies on high-quality tracks in order to form vertices. While the overall analysis strategy remains the same as in the earlier analysis, the improved techniques used here further reduce background and improve estimations of systematic uncertainties. For example, a new technique to suppress background vertices arising from accidental track intersections from additional proton-proton ($\Pp\Pp$) interaction vertices has been employed to reduce the number of background vertices by 40\%; the uncertainty due to the presence of \PQb quarks in the background template construction has been reduced from 41 to 6\%; and new procedures provide a more accurate evaluation of signal efficiencies and their corresponding systematic uncertainties. 

This analysis complements other searches for long-lived particles by the ATLAS and CMS experiments~\cite{Aaboud:2017iio,Aad:2020srt,Sirunyan:2020cao} in that it is highly sensitive to mean proper decay lengths between 0.1 and 15\mm. By requiring two reconstructed vertices inside the beam pipe, this search uniquely probes this region of parameter space using a set of stringent vertex and event selection criteria that results in a background-free search while retaining high signal efficiency for events containing pairs of long-lived particles.

The following sections address the CMS detector and event reconstruction (Section~\ref{sec:detector}), event samples and event preselection (Section~\ref{sec:samples}),  vertex reconstruction (Section~\ref{sec:vtxreco}),  the search strategy (Section~\ref{sec:strategy}), determination of the signal efficiency (Section~\ref{sec:sigeff}), construction of the background template (Section~\ref{sec:bkgest}), systematic uncertainties (Section~\ref{sec:syst}), results and statistical interpretation (Section~\ref{sec:results}), and, finally, a summary of our findings (Section~\ref{sec:summary}). An appendix provides a method for applying these results to other models that predict long-lived particles decaying to final states with two or more jets. Tabulated results are provided in HEPData~\cite{hepdata}.

\section{The CMS detector and event reconstruction}
\label{sec:detector}
The central feature of the CMS apparatus~\cite{Chatrchyan:2008zzk} is a superconducting solenoid of 6\unit{m} internal diameter, providing a magnetic field of 3.8\unit{T}. Within the solenoid volume are a silicon pixel and strip tracker, a lead tungstate crystal electromagnetic calorimeter (ECAL), and a brass and scintillator hadron calorimeter (HCAL), each composed of a barrel and two endcap sections. Forward calorimeters extend the pseudorapidity ($\eta$) coverage provided by the barrel and endcap detectors. Muons are detected in gas-ionization chambers embedded in the steel flux-return yoke outside the solenoid. Reference~\cite{Chatrchyan:2008zzk} provides a more detailed description of the CMS detector, together with a definition of the coordinate system used and the relevant kinematic variables. 

The vertex reconstruction at the crux of this analysis relies on the innermost detector surrounding the beam pipe, the upgraded silicon tracker. The tracker detects charged particles with $\abs{\eta} < 3$. Its innermost layer has a radius of 29\mm, and in total it has 1856 silicon pixel and 15\,148 silicon strip detector modules covering a total area of over 200~$\text{m}^2$, making it the largest silicon detector ever constructed. For nonisolated particles with transverse momentum \pt in the range $1 < \pt < 10\GeV$ and $\abs{\eta} < 3$,
 the track resolutions are typically 1.5\% in \pt and 20--75\mum in the transverse impact parameter (\dxy), defined as the distance of closest approach in the $x$-$y$ plane with respect to the center of the luminous region~\cite{CMS-DP-2020-049,Chatrchyan:2014fea}.  The \dxy resolution is approximately 25\% smaller than in earlier data sets, thanks to the silicon pixel tracker upgrade.

Events of interest are selected using a two-tiered trigger system~\cite{Khachatryan:2016bia}. The first level is composed of custom hardware processors used to select events at a rate of approximately 100~kHz, while the second level consists of a farm of processors running a version of the full event reconstruction software optimized for fast processing, and is used to reduce the event rate to about 1~kHz before data storage.

A particle-flow algorithm~\cite{CMS-PRF-14-001} aims to reconstruct and identify each individual particle in an event, with an optimized combination of information from the various elements of the CMS detector. The energy of photons is obtained from the ECAL measurement. The energy of electrons is determined from a combination of the electron momentum at the primary interaction vertex as determined by the tracker, the energy of the corresponding ECAL cluster, and the energy sum of all bremsstrahlung photons spatially compatible with originating from the electron track. The energy of muons is obtained from their \pt and polar angle. The energy of charged hadrons is determined from a combination of their momenta measured in the tracker and the matching ECAL and HCAL energy deposits, corrected for the response function of the calorimeters to hadronic showers. Finally, the energy of neutral hadrons is obtained from the corresponding corrected ECAL and HCAL energies.

Jets are reconstructed offline from particle-flow candidates clustered using the anti-\kt algorithm~\cite{Cacciari:2008gp, Cacciari:2011ma} with a distance parameter of 0.4. The jet momentum is determined as the vectorial sum of all particle momenta in the jet, and is found from simulation to be, on average, within 5 to 10\% of the true momentum over the whole \pt spectrum and detector acceptance. Additional proton-proton interactions within the same or nearby bunch crossings (pileup) can contribute extra tracks and calorimetric energy depositions, increasing the apparent jet momentum. To mitigate this effect, tracks identified to be originating from pileup vertices are discarded and an offset correction is applied to correct for remaining contributions. Jet energy corrections are derived from simulation studies so that the average measured response of jets becomes identical to that of particle-level jets. In situ measurements of the momentum balance in dijet, photon+jet, {\PZ}+jet, and multijet events are used to determine any residual differences between the jet energy scales in data and simulation, and appropriate corrections are made~\cite{Khachatryan:2016kdb}. We reject jets with parameters consistent with misidentified leptons, which may include misreconstructed electron or muon candidates~\cite{CMS-PAS-JME-16-003}. To identify jets originating from \PQb quark fragmentation (\PQb jets), the ``tight'' working point of the \textsc{DeepJet} tagging algorithm is used, which has an identification efficiency for \PQb jets from top quark decays with $\pt > 30\GeV$ of about 58\% and a misidentification probability for light-flavor jets (from the fragmentation of \PQu, \PQd, \PQs quarks and gluons) of about 0.1\%~\cite{Bols_2020,Sirunyan:2017ezt,CMS-DP-2018-058}.

Proton-proton interaction vertices are identified using high-quality tracks, and the one with the largest total physics-object $\pt^2$ is taken to be the primary $\Pp\Pp$ vertex, where the physics objects are the jets and missing transverse momentum associated with the vertex. The beam spot is identified with the mean position of the $\Pp\Pp$ interaction vertices. 

\section{Event samples and event preselection}
\label{sec:samples}
Events in both data and simulation are selected using a trigger requiring $\HT > 1050\GeV$, where \HT is the scalar sum of the jet \pt for jets with $\pt > 40\GeV$ and $\abs{\eta} < 2.5$. An offline requirement of $\HT > 1200$\GeV is imposed, and for both data and simulated events satisfying this requirement, the trigger efficiency is greater than 98\%. We also require at least four reconstructed jets, each with $\pt > 20$\GeV and $\abs{\eta}<2.5$. Together, these requirements define the event preselection criteria.

Signal events were simulated using \PYTHIA 8.230~\cite{PYTHIA} with the NNPDF3.1LO~\cite{Ball:2017nwa} set of parton distribution functions (PDFs), and the CP2 tune~\cite{Sirunyan:2019dfx} is used to model the underlying event. The samples are produced with ranges of masses (400--3000\GeV) and of $c\tau$ (0.1--100\mm). The event preselection efficiency is greater than 96\% for signal models where the mass of the long-lived particle is 1200\GeV or larger. For masses near 600\GeV, the event preselection efficiency is approximately 30\%--50\%.

Background events arising from SM processes that contain enough jet activity to satisfy the \HT trigger requirements come entirely from events with two or more jets produced through the strong interaction and events with pair-produced top quarks. These background samples are simulated using \MGvATNLO 2.4.2~\cite{MADGRAPH} with the NNPDF3.0~\cite{Ball:2014uwa} PDF set at leading order and with the MLM prescription~\cite{Alwall:2007fs} for combining matrix-element generators with parton showers. Simulation of the hadronization and showering is done with \PYTHIA 8.230~\cite{PYTHIA}, with the CP5~\cite{Sirunyan:2019dfx} tune.

Both background and signal samples use a \GEANTfour-based~\cite{GEANT} simulation for the CMS detector response. Simulated minimum-bias events are superimposed on the hard interaction in simulated events to match the observed pileup distribution in data.

\section{Vertex reconstruction}
\label{sec:vtxreco}
The displaced vertices are formed from charged-particle ``seed'' tracks. To ensure that reconstructed tracks are of high quality, we require tracks to satisfy several criteria: \pt of at least 1\GeV; at least one associated signal measured in the innermost layer of the pixel detector and at least one signal in an additional pixel layer; and signals measured in at least six layers of the silicon strip detector. These requirements result in a mean uncertainty in the transverse impact parameter of the tracks with respect to the beam spot of around 72\mum. Finally, the magnitude of the impact parameter in the transverse plane divided by its uncertainty, denoted \nsigmadxy, is required to be at least 4. This condition favors tracks with large impact parameters, thereby suppressing the SM background from tracks originating from the primary vertex. 

The next step in the vertex reconstruction procedure is to generate seed vertices from all pairs of seed tracks. The Kalman filter method~\cite{FRUHWIRTH1987444,BILLOIR1990219,BILLOIR1992139} is used to form a vertex from two or more tracks. The vertex is considered valid if its $\chi^2$ per degree of freedom is less than 5. If two vertices share a track and the three-dimensional distance between the vertex pair is less than 4 times the uncertainty in that distance, a vertex fit is applied to the complete set of tracks from both vertices. If the resulting fit satisfies the $\chi^2$ requirement, the two vertices are replaced by one single merged vertex. Otherwise, the two vertices remain separated, requiring a track arbitration step to decide which vertex is assigned the shared track. The track arbitration depends on the value of the track's three-dimensional impact parameter significance with respect to each of the vertices. If both impact parameters are within 1.5 standard deviations of both vertices, the shared track is assigned to the vertex with the larger number of tracks already; if the track has an impact parameter that is more than 5 standard deviations from either vertex, the shared track is removed from that vertex; otherwise, the shared track is assigned to the vertex to which it has the smaller impact parameter significance. When a track is dropped from a vertex, that vertex is refitted with its remaining tracks and replaced with a new vertex if the fit satisfies the $\chi^2$ requirement; otherwise the vertex is removed entirely. Pairs of vertices are merged iteratively following this algorithm until no two vertices share a track.

Occasionally, a vertex is formed from the accidental intersection of tracks that originate from separate pileup vertices. To suppress these, we consider each track associated with a vertex and calculate the shift in vertex position after refitting the vertex with that track removed. If the vertex position shifts by at least 50\mum along the beam axis, the track is permanently removed and the original vertex is replaced with this new refit vertex. This additional procedure is a new refinement in this analysis with respect to the previous CMS result and removes more than 40\% of background vertices in simulation with minimal impact on signal efficiency.

We select vertices with features consistent with a signal vertex by requiring two vertices to satisfy several criteria: at least five tracks; an $x$-$y$ displacement from the beam axis, defined as \dbv, of at least 100\mum to suppress background from displaced $\Pp\Pp$ interaction vertices; an $x$-$y$ position within a radius of 20.9\mm to suppress background vertices arising from interactions with material; and an uncertainty in \dbv of less than 25\mum to select vertices with a well-measured displacement and whose tracks have a large opening angle. This requirement on the \dbv uncertainty suppresses vertices from \PQb jets, which tend to have narrow opening angles between the associated tracks due to the large boost of \PQb hadrons relative to that of the massive particles in the signal. The efficiency of the signal vertex reconstruction and selection criteria is discussed in Section~\ref{sec:sigeff}.

Since the search focuses on signal models with pair-produced long-lived particles, we require that events have two vertices. Few events in the background contain one reconstructed displaced vertex; occurrences of higher vertex multiplicity events are even rarer. Simulations of background predict fewer than one event in the two-vertex search region for a data set corresponding to an integrated luminosity of \intlumiSecond. However, a reliable extraction of signal in data requires a more precise estimation of the background, which we evaluate using data as described in Section~\ref{sec:bkgest}.

While signal vertices must have at least five tracks in order to suppress background, vertices composed of three or four tracks function as useful control samples to validate the background estimation method. As seen in Table~\ref{tab:yields}, events with a single 3- or 4-track vertex are more common than events with a $\geq$5-track vertex by a factor of 30 or 6, respectively; moreover, the large background yield reduces the impact of any potential contamination by signal, so they provide a nearly pure background sample. As an example, for a multijet signal of mass 1600\GeV and mean proper decay length 10\mm, the expected signal contamination at the currently excluded cross section of 0.15\fb is below 0.1\%, with about one event in the 3-track one-vertex sample and about two events in the $\geq$5-track one-vertex sample.  Distributions of event-level variables (\eg, \HT, jet multiplicity) and vertex-level variables (\eg, \dbv, uncertainty in \dbv) are similar for background events with 3-, 4-, and $\geq$5-track vertices in both simulation and data. Therefore, vertices with lower track multiplicity provide a reliable sample for validation of the analysis procedure applied to the $\geq$5-track two-vertex sample.

\begin{table*}[htbp!]
\centering
\topcaption{Event yields in the control samples in data. The ``one-vertex'' events correspond to events containing exactly one vertex with the specified number of tracks. The ``two-vertex'' events have two or more vertices containing the specified numbers of tracks. We seek the signal in the $\geq$5-track two-vertex sample.}
\begin{scotch}{lcccc}
  Event category & 3-track & 4-track + 3-track & 4-track   & $\geq$5-track \\
\hline
One-vertex     & 61\,818 &                      \NA &   14\,730 &          2211 \\
Two-vertex     &     185 &                      101 &        12 &   See Section~\ref{sec:results} \\
\end{scotch}
\label{tab:yields}
\end{table*}

\section{Search strategy}
\label{sec:strategy}
\begin{figure}[hbtp]
\centering
\includegraphics[width=0.49\textwidth]{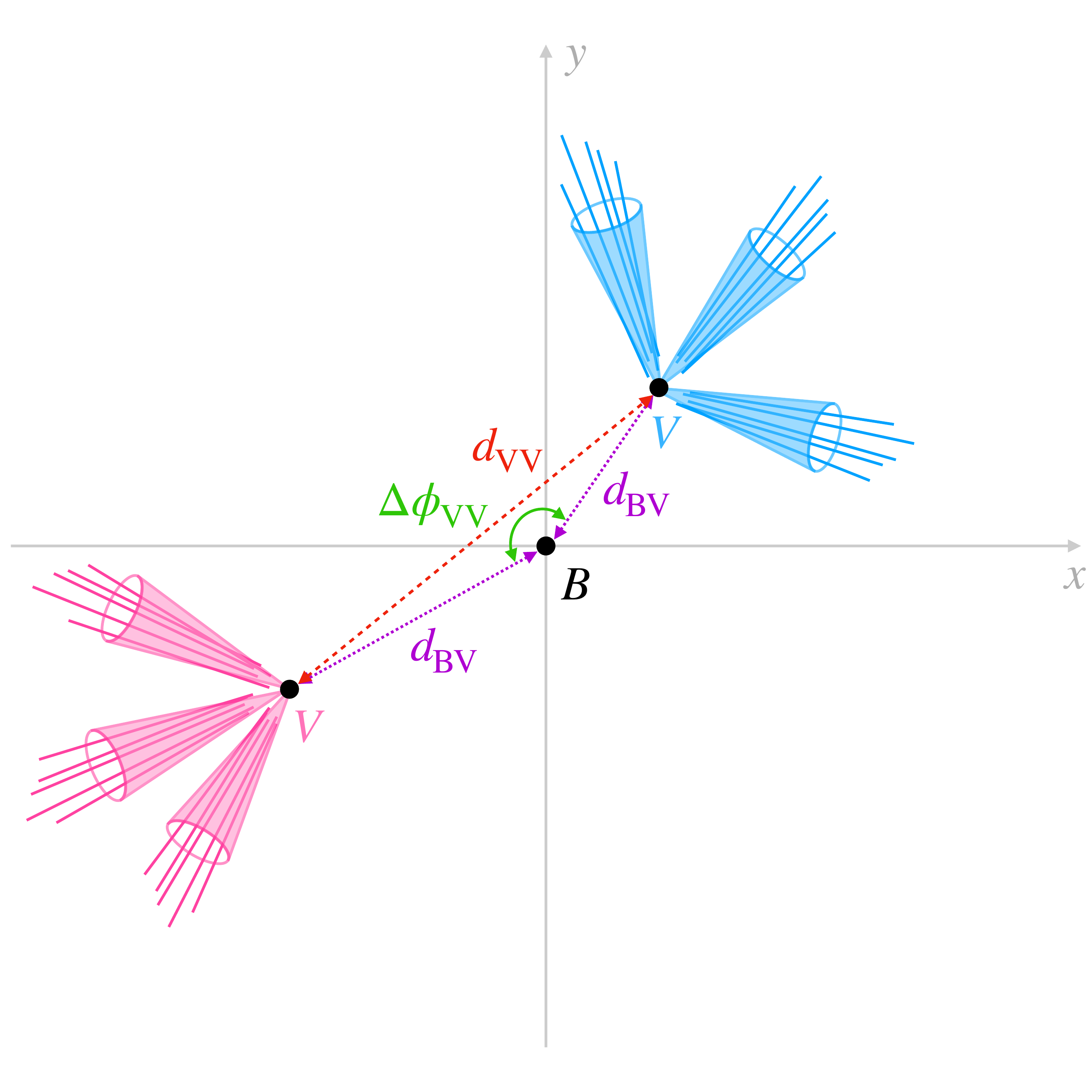}
\caption{Schematic diagram of an event with two signal vertices with the beam spot $B$ at the origin. The beam direction is perpendicular to the $x$-$y$ plane shown. The distance between the vertices is defined as \dvv. The distance from the beam spot to the vertices is defined as \dbv and the angle between the vertex displacement vectors is defined as \dphivv.}
\label{fig:eventcartoon}
\end{figure}

We select events that contain at least two vertices each with five or more tracks to search for pair-produced long-lived particles. We use the distance between two vertices in the $x$-$y$ plane, defined as \dvv and shown in Fig.~\ref{fig:eventcartoon}, as the discriminating variable between signal and the SM background. In signal events, the pair-produced long-lived particles tend to be emitted back-to-back in the $x$-$y$ plane, resulting in larger vertex separations than in the background, where \dvv tends to be small. In events with three or more vertices, the two vertices with the highest number of tracks are chosen for the \dvv calculation. If the number of tracks is equal, a mass value is assigned to the vertex, reconstructed from the momenta of the tracks associated with the vertex, and the one with the higher mass is chosen. However, in the 2017 and 2018 data, we observe no events with three or more vertices. 

\begin{figure}[hbtp]
\centering
\includegraphics[width=0.49\textwidth]{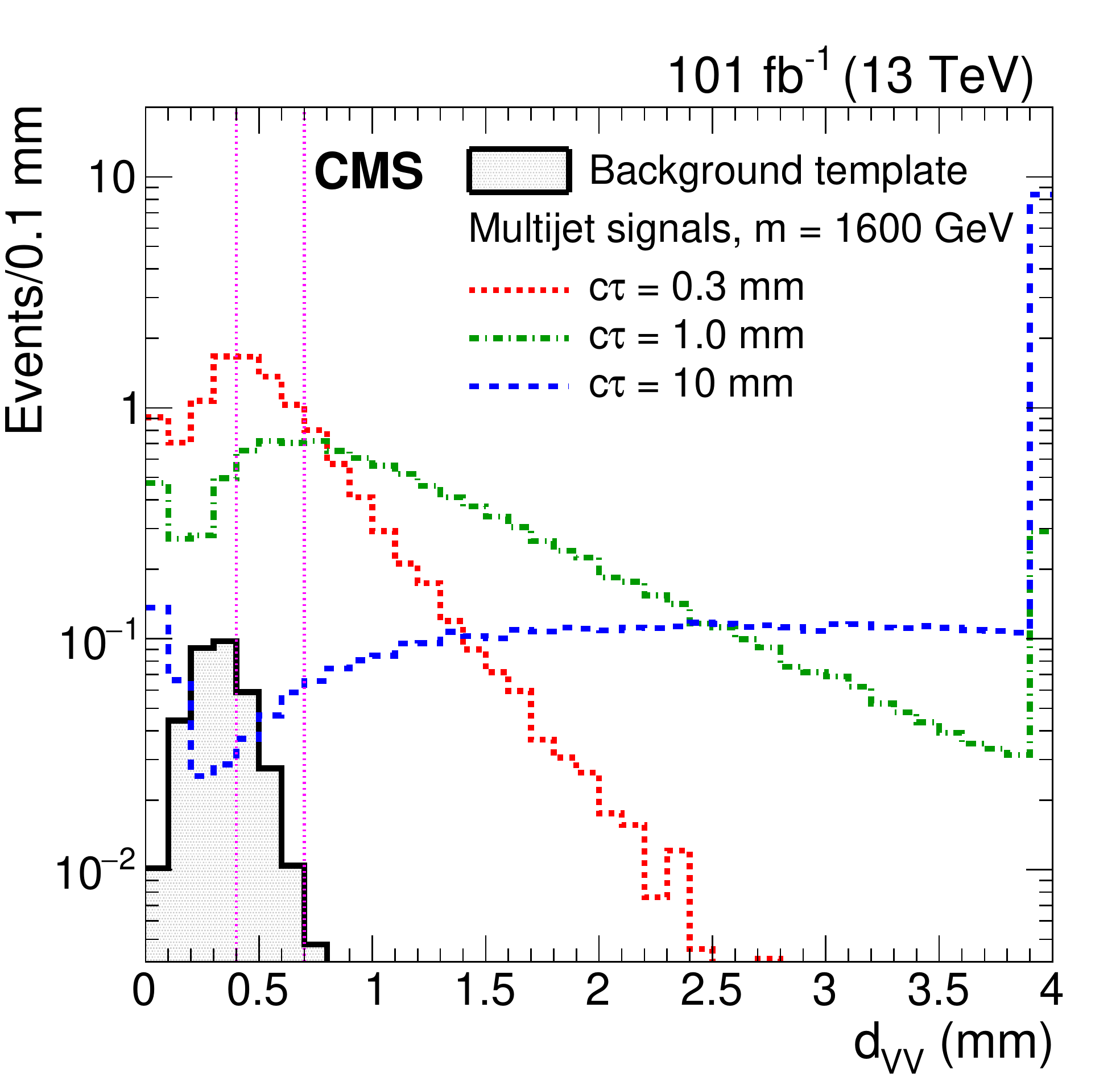}
\caption{The distribution of distances between vertices in the $x$-$y$ plane, \dvv, for three simulated multijet signals each with a mass of 1600\GeV, with the background template distribution overlaid. The production cross section for each signal model is assumed to be the lower limit excluded by Ref.~\cite{CMS-EXO-17-018}, corresponding to values of 0.8, 0.25, and 0.15\fb for the samples with $c\tau = 0.3$, 1.0, and 10\mm, respectively. The last bin includes the overflow events. The two vertical pink dashed lines separate the regions used in the fit.}
\label{fig:templates}
\end{figure}

The \dvv distribution of the background cannot be reliably ascertained from simulations. The simulated SM samples are smaller than the data samples, and background vertices are sensitive to the misreconstruction of tracks, which is difficult to accurately replicate in simulation. Thus we construct a \dvv background template using one-vertex events in data, as described in Section~\ref{sec:bkgest}, and validate the background estimate using the two-vertex control samples. Figure~\ref{fig:templates} compares the \dvv distributions for simulated multijet signals of various mean proper decay lengths and an LSP mass of 1600\GeV overlaid with the background template. The background peaks near 0.3\mm and has a 3\% probability of appearing above 0.7\mm, where the signal yield would be significant.

Ultimately, the background and signal templates are fit to the \dvv distribution observed in data to extract the signal yield. The fit uses three \dvv bins: 0--0.4, 0.4--0.7, and 0.7--40\mm. This binning scheme maximizes the signal significance in models with mean proper decay lengths in the 0.1 to 100\mm range.

\section{Signal efficiency}
\label{sec:sigeff}
To study the signal vertex reconstruction efficiency, we manually displace tracks from the primary vertex to produce artificial signal-like vertices in data and simulated SM events and then apply the reconstruction procedure. Starting from events with a well-recon\-structed primary vertex that satisfy the trigger and offline preselection requirements, we randomly select reconstructed light-flavor parton or \PQb quark jets that have $\pt > 50\GeV$ and at least four matched particle-flow candidate tracks. The jets are identified as light-flavor or \PQb quark jets based on whether or not they satisfy the \PQb-tagging criteria. The tracks associated with the selected jets are then displaced by a configurable distance in approximately the same direction as the vector sum of the selected jet momenta. The track impact parameter resolution in simulation is scaled to match data as a function of \pt and $\eta$. After track selection, vertex reconstruction, and vertex selection, we compute the fraction of events that contain a vertex reconstructed within 84\mum of the expected location, a condition that is satisfied by 95\% of vertices in signal simulation. This efficiency to reconstruct artificial vertices is comparable in simulated SM events to the vertex reconstruction efficiency in signal simulation, and reproduces the lifetime-dependence of the latter. The efficiency measured in data takes into account the track selection efficiency, which evolved with changes to the operating conditions such as temporary inefficiencies in the pixel detector~\cite{Trackergroup:2020cms}. Any loss of efficiency to reconstruct tracks with large displacements is reproduced in the simulation within a few percent based on a study of \PKzS mesons. This small discrepancy is incorporated into the systematic uncertainties described in Section~\ref{sec:syst}.

For the dijet signals, we replicate the signal by displacing the tracks associated with two light-flavor jets.  The efficiency is suppressed in events where the two jet momentum vectors are back-to-back and parallel to the displacement because of the large resulting uncertainty in vertex position. We therefore reweight the efficiency of these events relative to the others based on their relative proportions in the signal simulation. The differences in vertex reconstruction efficiency between data and simulation arise mainly from the modeling of the number of tracks satisfying the impact parameter requirement. These differences range from 5\% for the longest lifetimes to 16\% for the shortest.

For multijet signals, we displace the tracks associated with three light-flavor jets and two \PQb quark jets, which replicates the most common neutralino or gluino final state. Here, the reconstruction efficiency reaches 50\% in both data and simulation when six or seven seed tracks emanate from the vertex, and is over 90\% efficient for a vertex with at least 12 seed tracks. Vertices with large displacements typically have more seed tracks than those near the beam axis, because tracks with large displacements are more likely to pass the track impact parameter significance requirement. This leads to vertex reconstruction efficiencies of approximately 50\% (95\%) for samples with mean proper decay lengths of 100\mum (10\mm). On average, artificial vertices produced from data have two fewer seed tracks than those produced from simulated SM events, resulting in a lower vertex reconstruction efficiency in data for the short-lifetime samples. In the multijet case, after correcting for small differences in the track distributions of simulated events in the study and the signal, the vertex reconstruction efficiency difference between the data and simulation is between 0.1 and 14\%, with better agreement at longer lifetimes. For simulated multijet signals with LSP mass of 1600\GeV that satisfy the event preselection requirements and decay within the fiducial region considered, the efficiency to reconstruct two vertices in an event ranges from about 60 to 95\% for mean proper decay lengths of 0.3--30\mm, respectively.

The differences in the vertex reconstruction efficiencies between data and simulated SM events are used to correct the efficiency associated with each of the two displaced vertices in the signal events, resulting in a correction to the signal yield that primarily depends on the signal lifetime. A sample of these efficiency correction factors is provided in Appendix~\ref{app:reinterpretation}.

Figure~\ref{fig:sigeff} shows the signal efficiency in both multijet and dijet signals after applying all event and vertex requirements. The increase in the efficiency with mass comes from the higher probability of satisfying the trigger and offline \HT requirements, in addition to higher track multiplicity from the decay vertex. At small $c\tau$, the efficiency increases with lifetime while moving away from the prompt region, but decreases for large lifetimes because of the requirement that vertices lie within the beam pipe.

\begin{figure}[hbtp]
\centering
\includegraphics[width=0.49\textwidth]{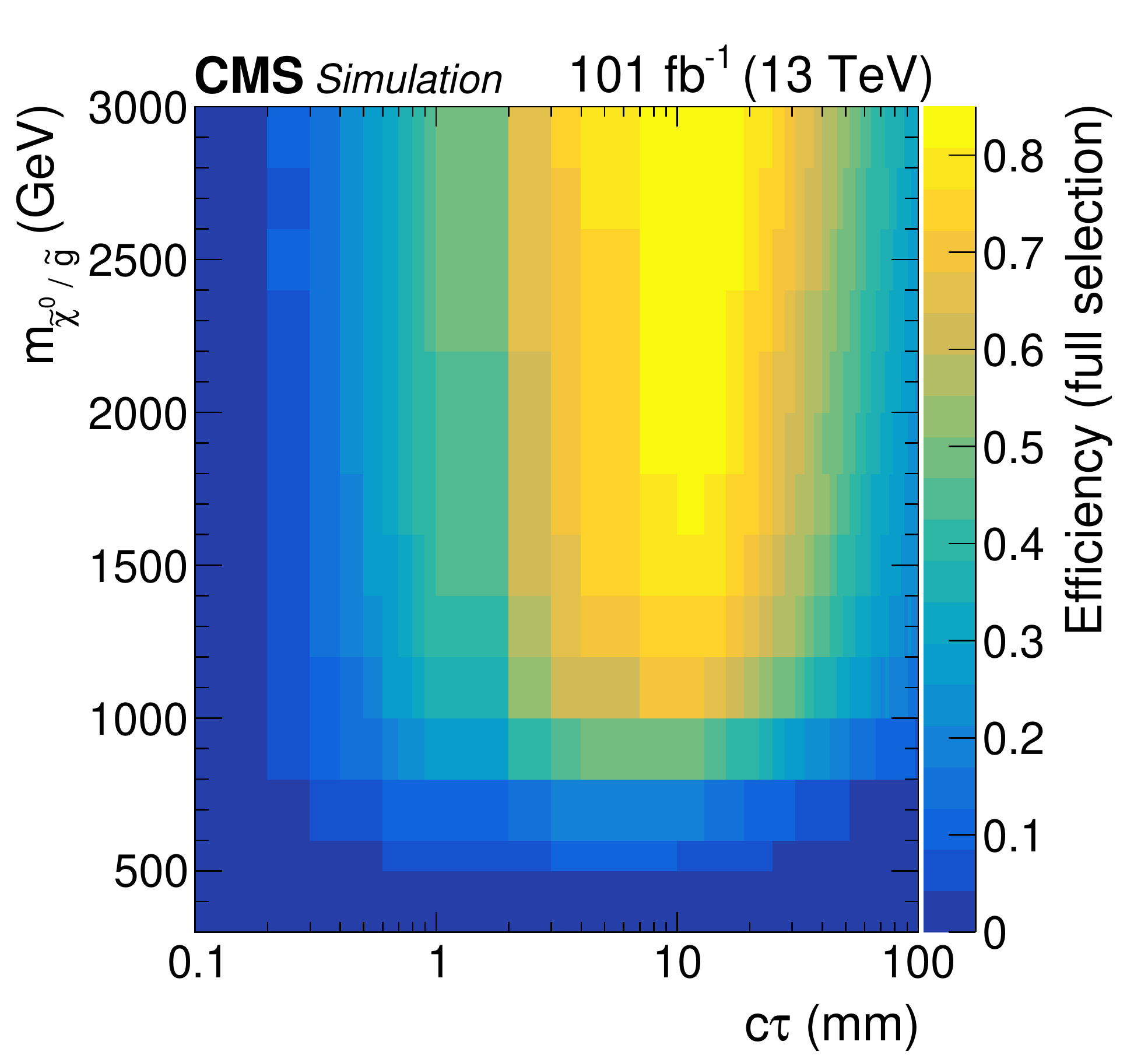}
\includegraphics[width=0.49\textwidth]{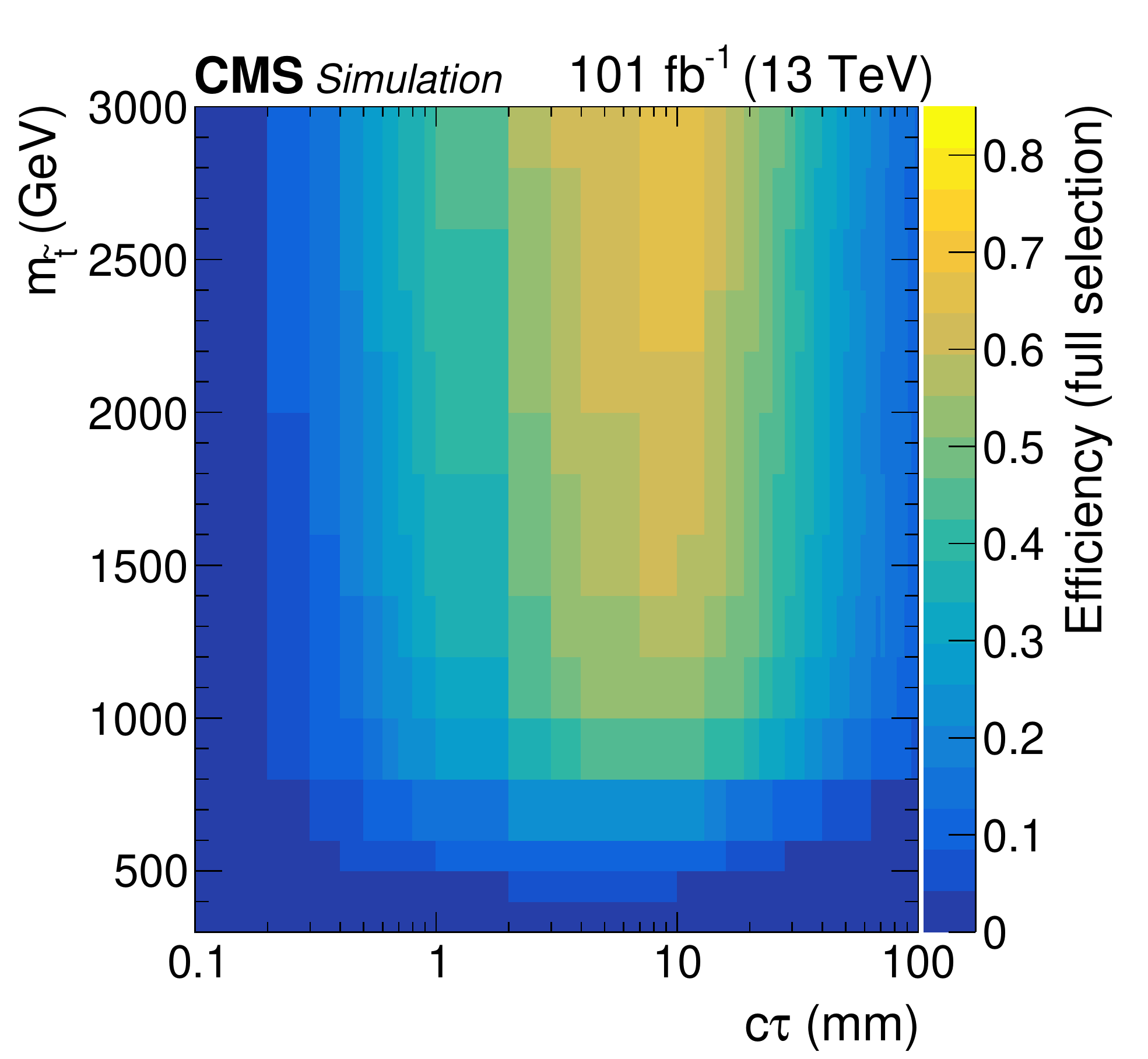}
\caption{Multijet (\cmsLeft) and dijet (\cmsRight) signal efficiencies as a function of the signal mass and lifetime for events satisfying all event and vertex requirements, with corrections based on systematic differences in the vertex reconstruction efficiency between data and simulation.}
\label{fig:sigeff}
\end{figure}

\section{Background template}
\label{sec:bkgest}
In the background, most displaced vertices are spurious and include at least one poorly reconstructed track that endows the vertex with a displacement away from the interaction point. Individual well-reconstructed \PQb jets do not contribute a significant number of background vertices because of the stringent requirement on the vertex \dbv uncertainty. As a result, the primary background to this search comes from events containing two spurious displaced vertices. The displacements of these vertices are independent of one another, except for correlations due to events with \PQb quarks. These correlations are handled with separate treatments of events with and without \PQb-tagged jets. The independence of the two vertex displacements is a crucial feature, as it offers a method to predict the shape of the search variable distribution, \dvv, in two-vertex events by using information from events containing only one vertex. The constructed template, denoted as \dvvc, provides the predicted two-vertex yields in each of the three \dvv search bins. Events with one vertex are more common than two-vertex events by a factor of 100 to 1000, as shown in Table~\ref{tab:yields}. This abundance of one-vertex events is used to create a template with high statistical precision.

A single value of \dvvc is constructed from two values of \dbv randomly chosen from the \dbv distribution in one-vertex events, along with a random value of \dphivv, which specifies the azimuthal angle between the vertex displacement vectors, as shown in Fig.~\ref{fig:eventcartoon}. The sampling repeats until the number of entries in the \dvvc template is 20 times the number of one-vertex events in data. The large sample size reduces the statistical uncertainty and increases the probability of adequately probing the tail of the \dbv distribution; further sampling provides no additional reduction to the statistical uncertainty. The details of the input variables to the \dvvc template, along with corrections, are described in the following paragraphs.

\begin{figure}[hbtp]
\centering
\includegraphics[width=0.49\textwidth]{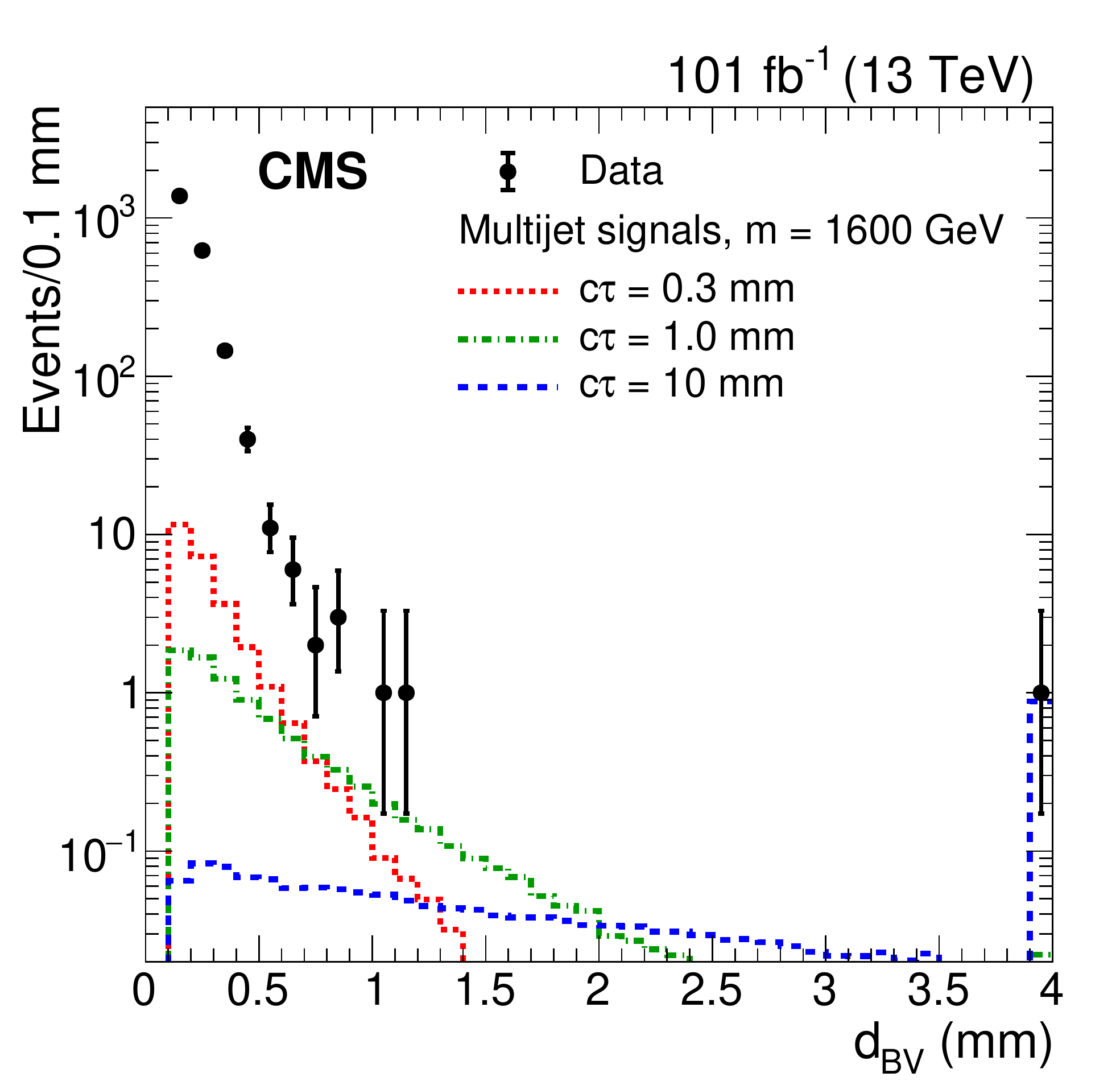}
\caption{The distribution of \dbv for $\geq$5-track one-vertex events in data and three simulated multijet signal samples each with a mass of 1600\GeV. The production cross section for each signal model is assumed to be the lower limit excluded by Ref.~\cite{CMS-EXO-17-018}, corresponding to values of 0.8, 0.25, and 0.15\fb for the samples with $c\tau = 0.3$, 1.0, and 10\mm, respectively. The last bin includes the overflow events. This bin includes one event in data with a vertex with large \dbv that appears to arise from tracks originating from separate $\Pp\Pp$ interaction vertices, consistent with background.}
\label{fig:dbv}
\end{figure}

The distributions of \dbv in $\geq$5-track one-vertex events for data and for simulated signal samples of varying lifetimes are shown in Fig.~\ref{fig:dbv}. The effects of signal contamination at the maximum level consistent with existing upper limits on the signal cross sections are negligible because of the much larger one-vertex background at low \dbv.

The distribution of azimuthal angles between all possible pairs of jets in an event, denoted as \dphijj, has a preference for high-angle separations and roughly corresponds to the distribution of \dphivv for events with two low-track-multiplicity vertices. Since the \dphijj distribution is consistent across events containing vertices with different track multiplicities, the \dphijj distribution for the large sample of 3-track one-vertex events is used to sample a \dphivv angle for the \dvvc template construction.

The vertex reconstruction procedure merges nearby vertices, suppressing small values of \dvv. To capture this behavior in the template, we correct the \dvvc template using the survival efficiency of vertex pairs as a function of their separation. This efficiency is estimated in data from the fraction of initial 3-track vertex pairs that remain after merging.

Single \PQb jet vertices rarely satisfy the requirement on the \dbv uncertainty because the narrow collimation of tracks from the \PQb jet results in poor \dbv resolution. However, events with \PQb quarks are four times more likely to have a displaced vertex than those without because the \PQb jet tracks are more likely to satisfy the \nsigmadxy requirement. Moreover, in simulated background samples, \PQb quark events are observed to have vertices with larger \dbv on average by 10\%--20\%. Thus, events with \PQb quark pairs introduce correlations in the \dbv of vertex pairs that are not captured in the template construction, which pairs vertices associated with light-flavor or \PQb quark jets at random. This effect is handled by constructing separate \dvvc templates for events with and without a \PQb-tagged jet. These templates are combined into a single template by weighting them according to the expected fraction of two-vertex events with and without \PQb quarks. The percentage of \PQb quark events is determined in simulation by using the \PQb jet identification efficiencies and misidentification probabilities (58\% and 0.1\% on average, respectively), along with their corresponding data-to-simulation correction factors, to relate \PQb-tagged events to \PQb quark events. The percentages of \PQb quark events are 85, 89, and 95\% in 3-, 4-, and $\geq$5-track two-vertex events, respectively. This procedure leads to a 53\% enhancement in the yield in the third \dvvc bin (0.7--40\mm).

For the normalizations of the signal and background templates, we extract the signal yield from a fit to the observed \dvv distribution. In the background-only fit, which is also used to fit data in the control samples, the template is normalized to the total two-vertex event yield observed in data, thus providing a background prediction in each of the three \dvv bins. In situations in which no two-vertex events are observed, the template is normalized using a procedure similar to that used to construct the template shape; this relies on information from one-vertex events and the assumption that displaced vertices in background events are independent of one another. The normalization is then calculated by combining the trigger and preselection efficiencies with the squared vertex reconstruction efficiency for events with and without \PQb quarks, corrected for the survival efficiency of nearby vertex pairs. We validate this procedure using the control samples, where the ratios of the observed yields to the predicted yields are $1.02^{+0.08}_{-0.07}$, $1.11^{+0.12}_{-0.11}$, and $1.05^{+0.40}_{-0.30}$ for events with two 3-track vertices, exactly one 4- and one 3-track vertex, and two 4-track vertices, respectively.

\begin{figure*}[hbtp]
\centering
\includegraphics[width=0.49\textwidth]{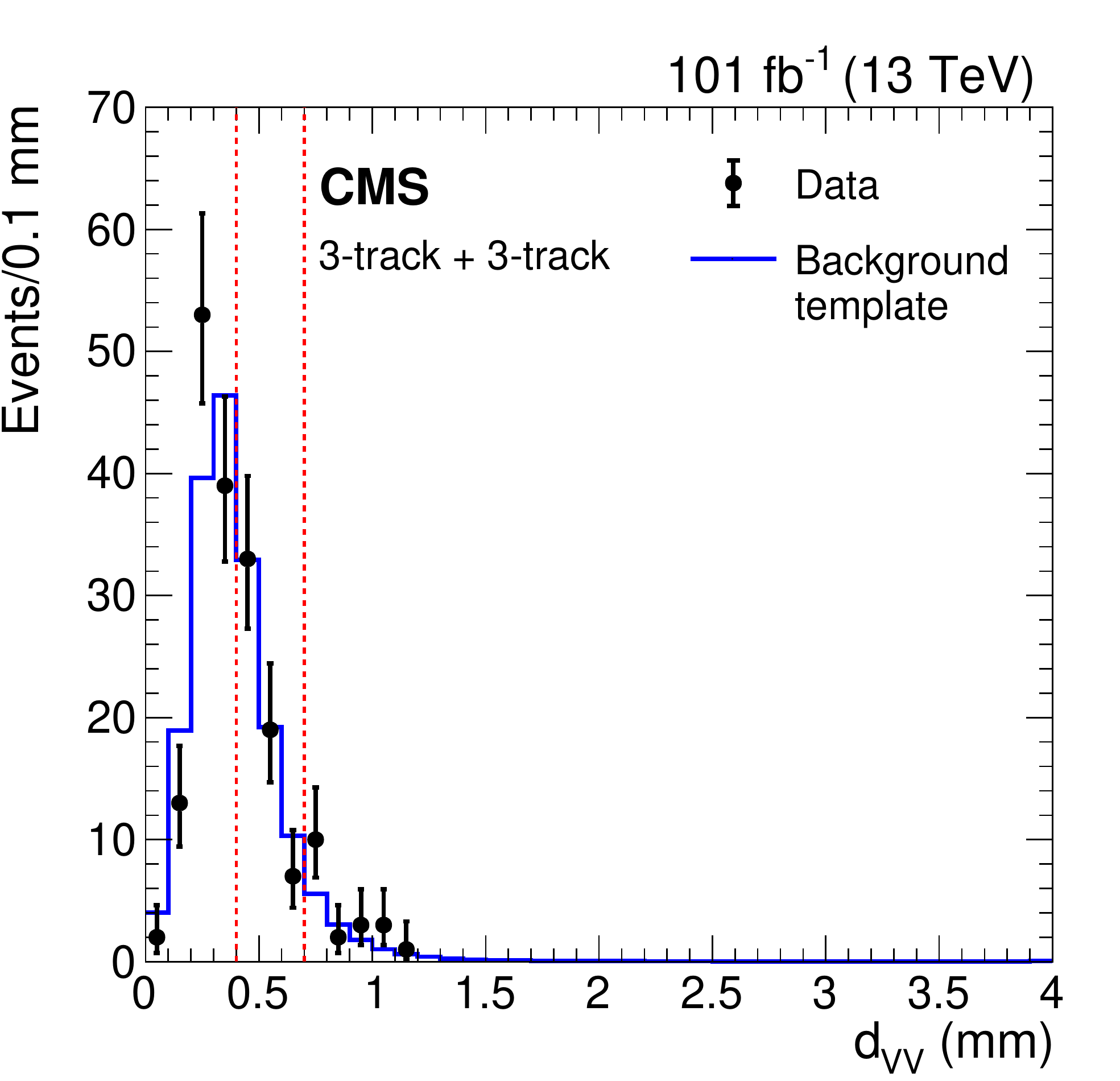}
\includegraphics[width=0.49\textwidth]{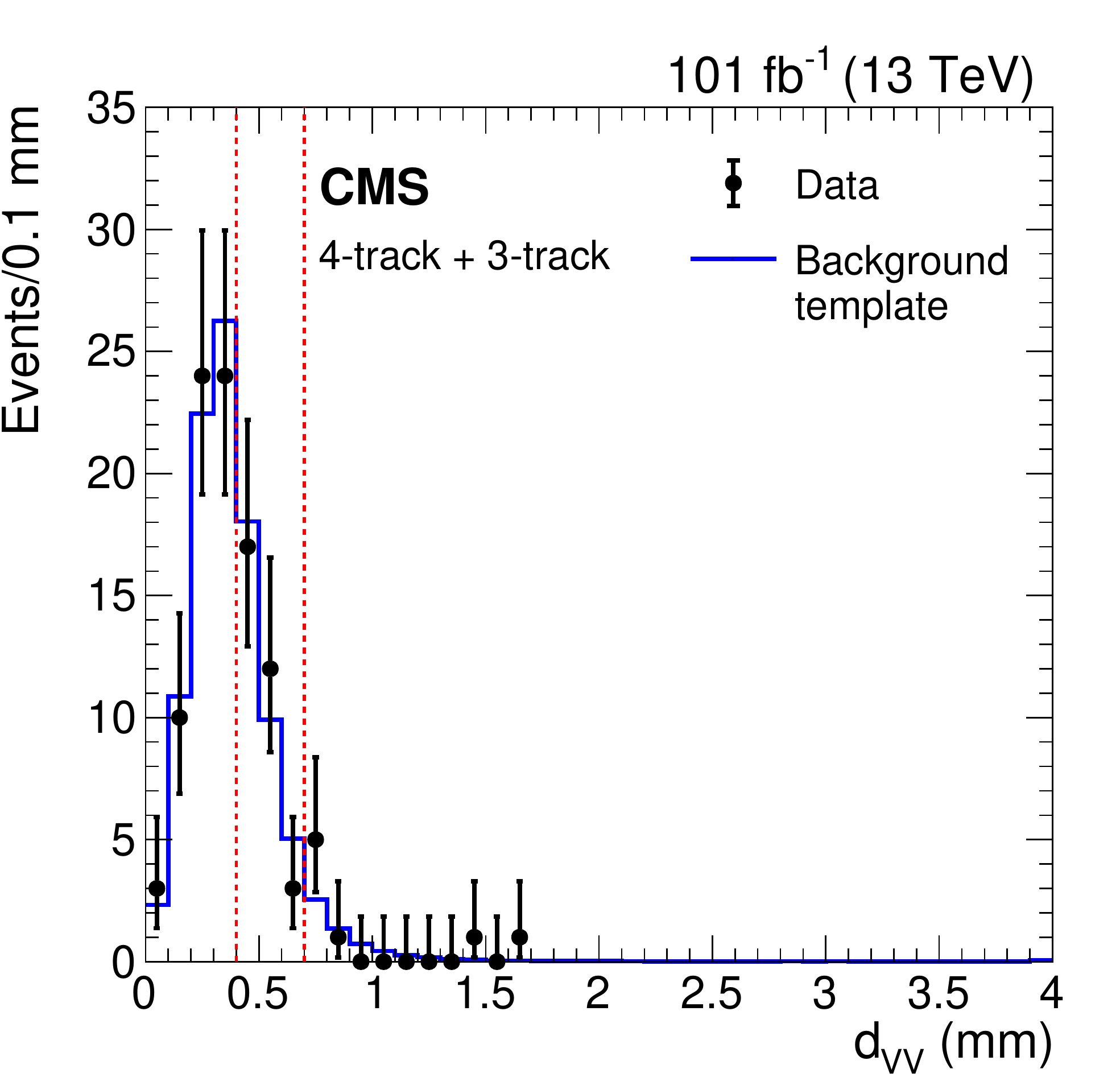}
\includegraphics[width=0.49\textwidth]{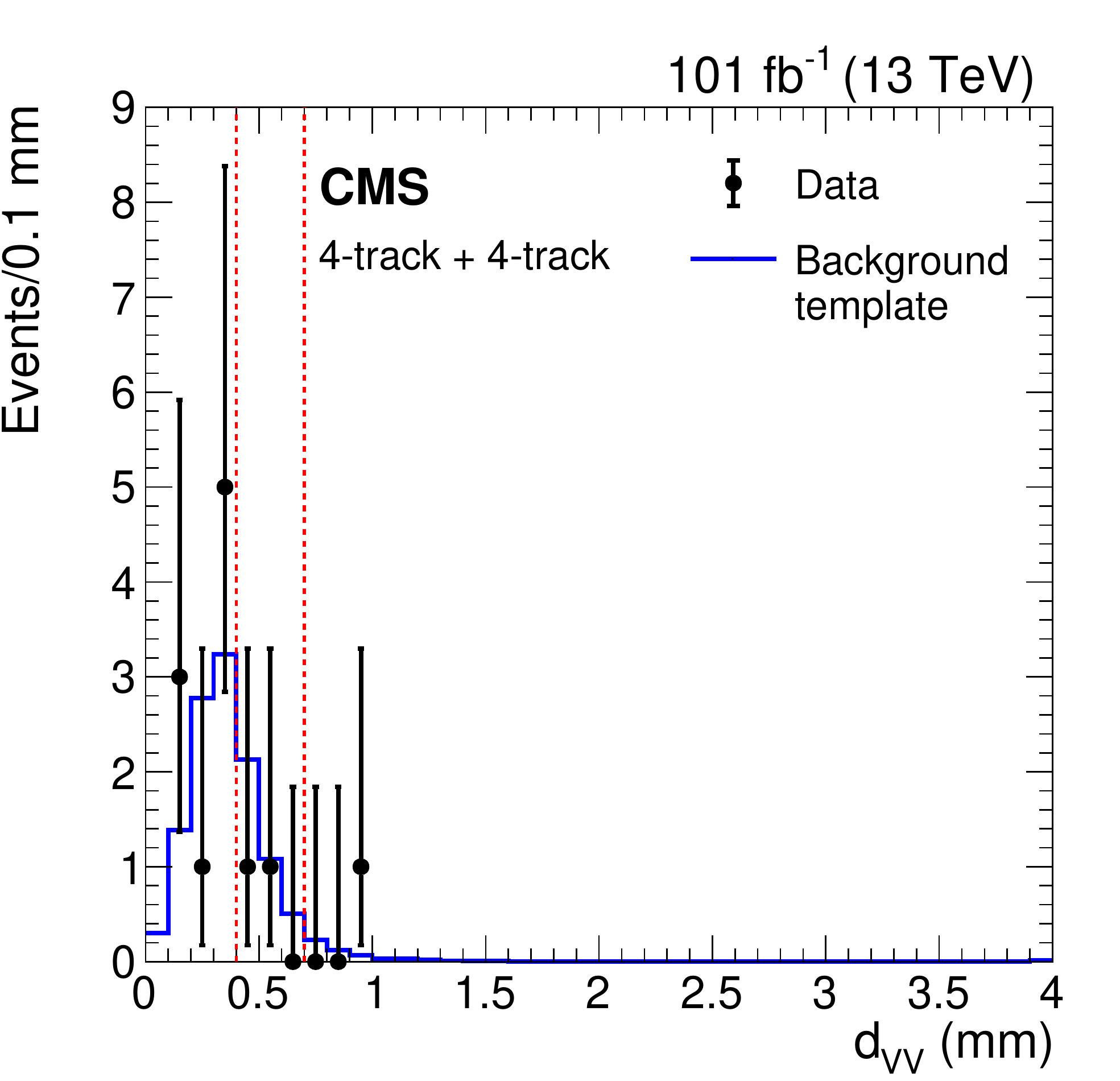}
\includegraphics[width=0.49\textwidth]{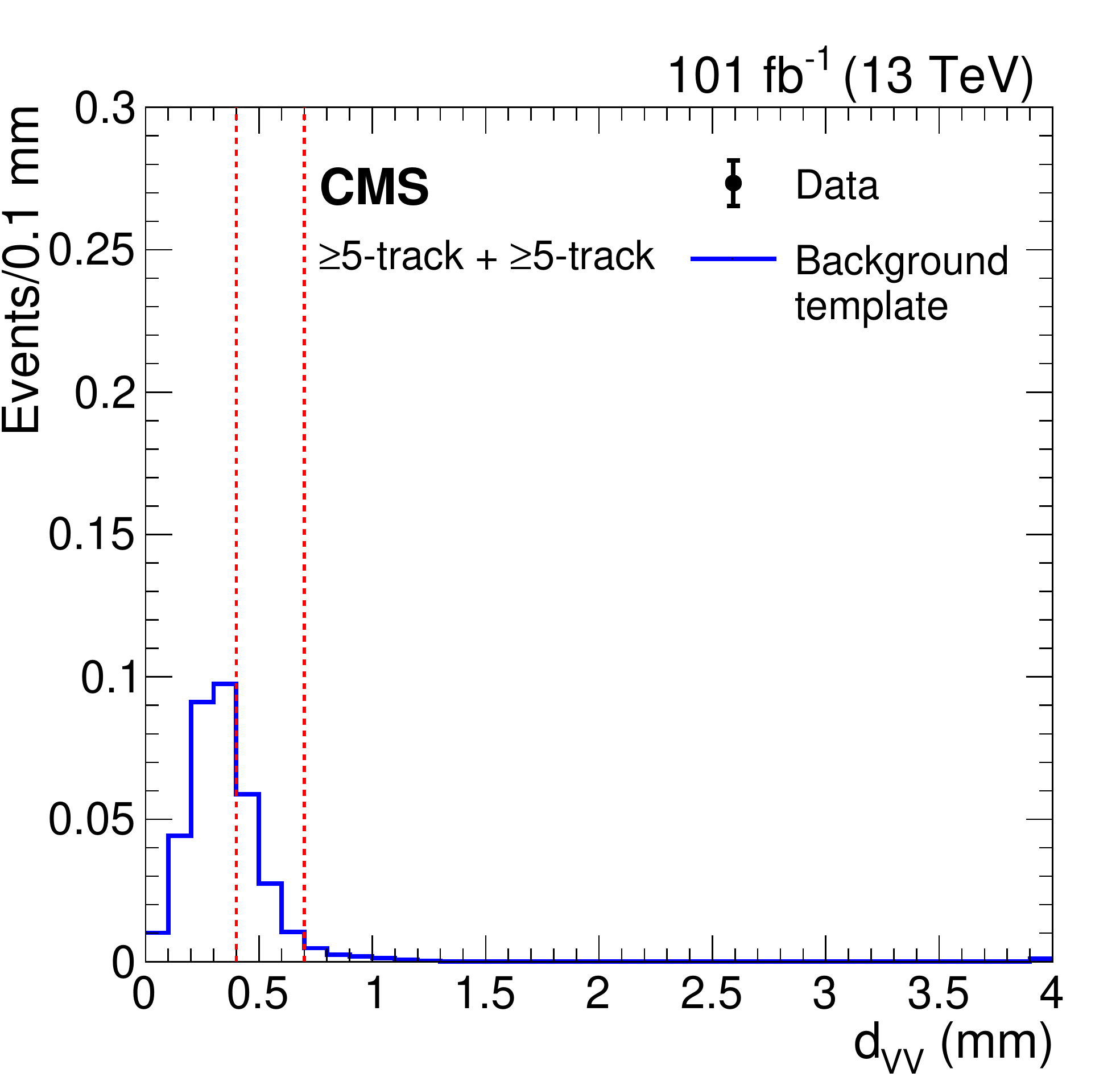}
\caption{Distribution of the $x$-$y$ distances between vertices, \dvv, for 2017 and 2018 data. The background distribution \dvvc (blue continuous line) is constructed from one-vertex events in data, and is normalized to the number of two-vertex events in data with two 3-track vertices (upper left), events which have exactly one 4-track vertex and one 3-track vertex (upper right), and events with two 4-track vertices (lower left). The background distribution \dvvc for $\geq$5-track two-vertex events (lower right) is normalized using one-vertex event information. The two vertical red dashed lines separate the regions used in the fit.}
\label{fig:closure}
\end{figure*}

Figure~\ref{fig:closure} compares the background templates to the observed two-vertex \dvv distributions. In the control samples, the yields in each of the three \dvvc bins in data are consistent with predictions from the template. The results in the signal region are discussed in Section~\ref{sec:results}.

\section{Systematic uncertainties}
\label{sec:syst}
\subsection{Systematic uncertainties in signal reconstruction}
Since the fit uses signal \dvv templates from simulation, potential differences between data and simulation give rise to systematic uncertainties. The dominant uncertainty comes from the vertex reconstruction efficiency, with other effects such as the PDF and strong coupling constant \alpS uncertainties in the simulation, pileup, jet energy resolution and scale, integrated luminosity, trigger efficiency, and run conditions providing smaller contributions. 

We assign a systematic uncertainty associated with the correction to the signal vertex reconstruction efficiency described in Section~\ref{sec:sigeff} based on the following considerations: sensitivity to the number and flavor of selected jets, the distribution of vertex displacements, the fraction of selected jets with back-to-back momentum vectors for dijet signals, and other smaller effects. This systematic uncertainty, which is always equal to or larger than the size of the correction, falls within the ranges of 11\%--41\% for dijet signals and 1\%--36\% for multijet signals, with the largest uncertainties at low masses and short mean proper decay lengths.

The remaining systematic uncertainties related to the signal efficiency are much smaller. The impact of the PDF uncertainty is estimated using simulation samples generated with reweighted NNPDF replica sets~\cite{Butterworth:2015oua} and the parton shower \alpS uncertainty is evaluated using variations of the renormalization scale and nonsingular terms, separately for initial- and final-state radiation~\cite{Mrenna:2016sih}. Together, the PDF and \alpS uncertainties yield a range of uncertainty between 1 and 8\%, depending primarily on the signal mass because of the underlying uncertainty in the parton luminosities, which varies with the particle mass~\cite{Ball:2017nwa}. Variations to the renormalization and factorization scales at the matrix-element level have a negligible impact on the signal efficiency.

The uncertainty in the integrated luminosity is 2.3\% in 2017~\cite{CMS-PAS-LUM-17-004} and 2.5\% in 2018~\cite{CMS-PAS-LUM-18-002}. Uncertainties in the jet energy scale can affect the probability of satisfying the offline \HT and jet \pt requirements. Variations of the jet energy scale result in changes in the signal efficiency of 5\% or less for all signal sample masses and lifetimes. Similarly, variations of the jet energy resolution result in differences of 2\% or less in the signal efficiency. The uncertainty in the signal efficiency due to pileup is 2\%. The trigger efficiency differences between data and simulation contribute an uncertainty of 1\%. Certain run conditions during the data collection led to inefficiencies in the electromagnetic and hadronic calorimeters affecting jets in parts of the detector, ultimately contributing 1\% uncertainty for each separate issue.

Table~\ref{tab:sigeffsyst} summarizes the systematic uncertainties related to the signal models. We assume no correlations between the different contributions and obtain an overall systematic uncertainty by adding each value in quadrature.

\begin{table*}[htbp!]
\centering
\topcaption{Signal-related systematic uncertainties for dijet and multijet signal models.  The total uncertainty is the sum in quadrature of the individual components. The ranges presented reflect differences among the various signal mass and lifetime hypotheses, as well as differences between the 2017 and 2018 data.}
\begin{scotch}{lcc}
Systematic effect      & Dijet uncertainty (\%) & Multijet uncertainty (\%) \\
\hline
Vertex reconstruction     & 11--41 & 1--36 \\
PDF and \alpS uncertainty & 1--8   & 1--8 \\
Integrated luminosity     & 2--3   & 2--3 \\
Jet energy scale          & 5      & 5 \\
Jet energy resolution     & 2      & 2 \\
Pileup                    & 2      & 2 \\
Trigger efficiency        & 1      & 1 \\
Changes in run conditions & 1      & 1 \\[\cmsTabSkip]
Total                     & 13--42 & 7--36 \\
\end{scotch}
\label{tab:sigeffsyst}
\end{table*}

\subsection{Systematic uncertainties in the background template}
Systematic uncertainties in the background template come from effects that modify the shape of the constructed \dvvc distribution away from the shape of the true two-vertex \dvv distribution. The 3-track vertex control sample provides a statistically precise way to assess these differences. Thus, within each of the three bins in the \dvvc template in the 3-track vertex control sample, we evaluate the ratio of the yield predicted by the template to the observed 3-track two-vertex yield in data, referred to as the closure. The deviation from unity, added in quadrature with its statistical uncertainty, is used as a measure of the systematic uncertainty for each \dvvc bin. We find that the \dvv/\dvvc ratios are $0.99 \pm 0.10$ in the 0--0.4\mm bin, $0.93 \pm 0.12$ in the 0.4--0.7\mm bin, and $1.38 \pm 0.32$ in the 0.7--40\mm bin. 

The normalization of the background template is calculated following the same principle as the template itself. Thus, the same variations are taken to assess the systematic uncertainty in the normalization factor. The dominant contributor driving the size of this uncertainty is the vertex pair survival efficiency correction. This systematic uncertainty is assigned equally to all three bins.

The assumption that the closure in 3-track events implies closure in $\geq$5-track events is tested with variations of the different inputs and corrections to the template. The template shape is particularly sensitive to the vertex pair survival efficiency correction, which uses the \dvv-dependent efficiency for vertex pairs to survive. To vary this procedure and derive an alternative efficiency curve, we consider seed vertices formed from all possible combinations of five tracks. We construct the \dvvc template with the resulting efficiency curve using this variation and assign the relative difference per bin of the template as the systematic uncertainty.

When constructing the background template, the angular separation between vertices \dphivv is modeled from the \dphijj distribution in 3-track vertices. The \dphijj distributions in $\geq$5-track one-vertex events and 3-track one-vertex events are consistent, but this does not exclude differences in the angles between jets and vertices. To gauge this effect, we construct the template by sampling the \dphivv value from a uniform distribution. The relative difference of the resulting template from the nominal template in each \dvvc bin is taken as the systematic uncertainty.

The \PQb tagging efficiencies and misidentification probabilities are determined using simulated events in the phase space relevant to this analysis, and efficiency correction factors are applied to match those observed in data. We vary these corrections within the limits allowed by measurements of the \pt-dependent \PQb tagging efficiency~\cite{Sirunyan:2017ezt} and take the relative difference between the resulting templates as the systematic uncertainty. Similarly, we vary the \PQb quark fraction in $\geq$5-track vertex events within the ranges observed in 3- and 4-track vertex events, assigning the systematic uncertainty as the relative difference in the resulting template.

Table~\ref{tab:bkgestsystdata} summarizes the systematic uncertainty for each of these components for each \dvv bin. We assume no correlations between these different effects and add all values in quadrature to obtain the overall systematic uncertainty in each bin. To preserve the normalization of the template, the limits are computed assuming that the first bin, which contains most of the background events, has a background systematic uncertainty that is anti-correlated with those in the second and third bins. Additionally, each bin is fully correlated across the different data taking periods, with the statistical components of each bin assumed to be uncorrelated.

\begin{table*}[htbp!]
\centering
\topcaption{Systematic uncertainties in the background prediction in each \dvvc bin arising from varying the construction of the \dvvc
template.  The total systematic uncertainty in each bin is the sum in quadrature of the values, assuming no correlations among the sources.}
\cmsTable{
\begin{scotch}{lccc}
                                                            & \multicolumn{3}{c}{Systematic uncertainty (\%)} \\
Systematic effect                                           & 0--0.4\mm       & 0.4--0.7\mm     & 0.7--40\mm  \\
\hline                                                                                                                
Closure in 3-track control sample                           & $10$            & $14$            & $50$   \\
$\geq$5-track template normalization factor                 & $24$            & $24$            & $24$   \\
Difference from 3-track vertices to $\geq$5-track vertices: &                 &                 &        \\
\hspace{0.1in} Modeling of vertex pair survival efficiency  & $9$             & $20$            & $25$   \\
\hspace{0.1in} Modeling of \dphivv                          & $3$             & $6$             & $6$    \\
\hspace{0.1in} Variation of \PQb quark fraction          & $1$             & $3$             & $6$    \\
Variation of \PQb tagging correction factors             & $0.5$           & $0.5$           & $1$    \\[\cmsTabSkip]
Total                                                       & $28$            & $35$            & $61$   \\
\end{scotch}
}
\label{tab:bkgestsystdata}
\end{table*}

\section{Results and statistical interpretation}
\label{sec:results}
Table~\ref{tab:bincounts} summarizes the predicted $\geq$5-track two-vertex event yields in each of the three \dvv bins from the background and signal templates for three multijet signal lifetime points, as well as the observation in data. No $\geq$5-track two-vertex events were observed in the 2017 and 2018 data. 

\begin{table*}[hbtp!]
\centering
\topcaption{Predicted yields for the background-only normalized template, predicted yields for three simulated multijet signals each with a mass of 1600\GeV, and the observed yield in each \dvv bin. The production cross section for each signal model is assumed to be the lower limit excluded by Ref.~\cite{CMS-EXO-17-018}, corresponding to values of 0.8, 0.25, and 0.15\fb for samples with $c\tau = 0.3$, 1.0, and 10\mm, respectively. The uncertainties in the signal yields and the systematic uncertainties in the background prediction reflect the systematic uncertainties given in Tables~\ref{tab:sigeffsyst} and~\ref{tab:bkgestsystdata}, respectively.}
\cmsTable{
\begin{scotch}{lcrccc}
             &                                       & \multicolumn{3}{c}{Predicted multijet signal yields} &  \\
$\dvv$ range & Predicted background yield            & {0.3\mm} & {1.0\mm} & {10\mm} & Observed \\
\hline
0--0.4\mm    & $0.243 \pm 0.003\stat \pm 0.061\syst$ & $4.4 \pm 0.5$ & $ 1.5 \pm 0.1$ & $ 0.26 \pm 0.02$  & 0 \\
0.4--0.7\mm  & $0.097 \pm 0.003\stat \pm 0.032\syst$ & $4.1 \pm 0.5$ & $ 2.1 \pm 0.2$ & $ 0.14 \pm 0.01$  & 0 \\
0.7--40\mm   & $0.012 \pm 0.001\stat \pm 0.006\syst$ & $3.0 \pm 0.3$ & $ 7.6 \pm 0.7$ & $12 \pm 1  $   & 0 \\
\end{scotch}
}
\label{tab:bincounts}
\end{table*}

To extract the signal yield from the data, we perform a binned shape fit using an extended maximum likelihood with three \dvv bins. Signal \dvv templates come directly from simulation with a template for each signal model, mass, and lifetime point. The background \dvvc template is constructed from the one-vertex events in data. The overall normalizations of the signal and background templates are free parameters of the fit under the constraint that the signal yield is not negative. The results obtained from the fit depend on the relative yields in the three \dvv bins and their systematic uncertainties. The 2017 and 2018 data sets are treated independently and combined in the fit.

The upper limits on the signal cross section are determined by first assuming a uniform Bayesian prior for the cross section. For each signal mass and lifetime point, the signal efficiency is constrained by a log-normal prior with a corresponding width as determined from the overall systematic uncertainty in the signal processes as summarized in Table~\ref{tab:sigeffsyst}. The shape uncertainty in the signal template arises from the statistical uncertainty in the simulation. For the background template, a log-normal prior is taken for each \dvvc bin for each year with widths specified in Table~\ref{tab:bkgestsystdata}. 

\begin{figure*}[hbtp]
\centering
\includegraphics[width=0.49\textwidth]{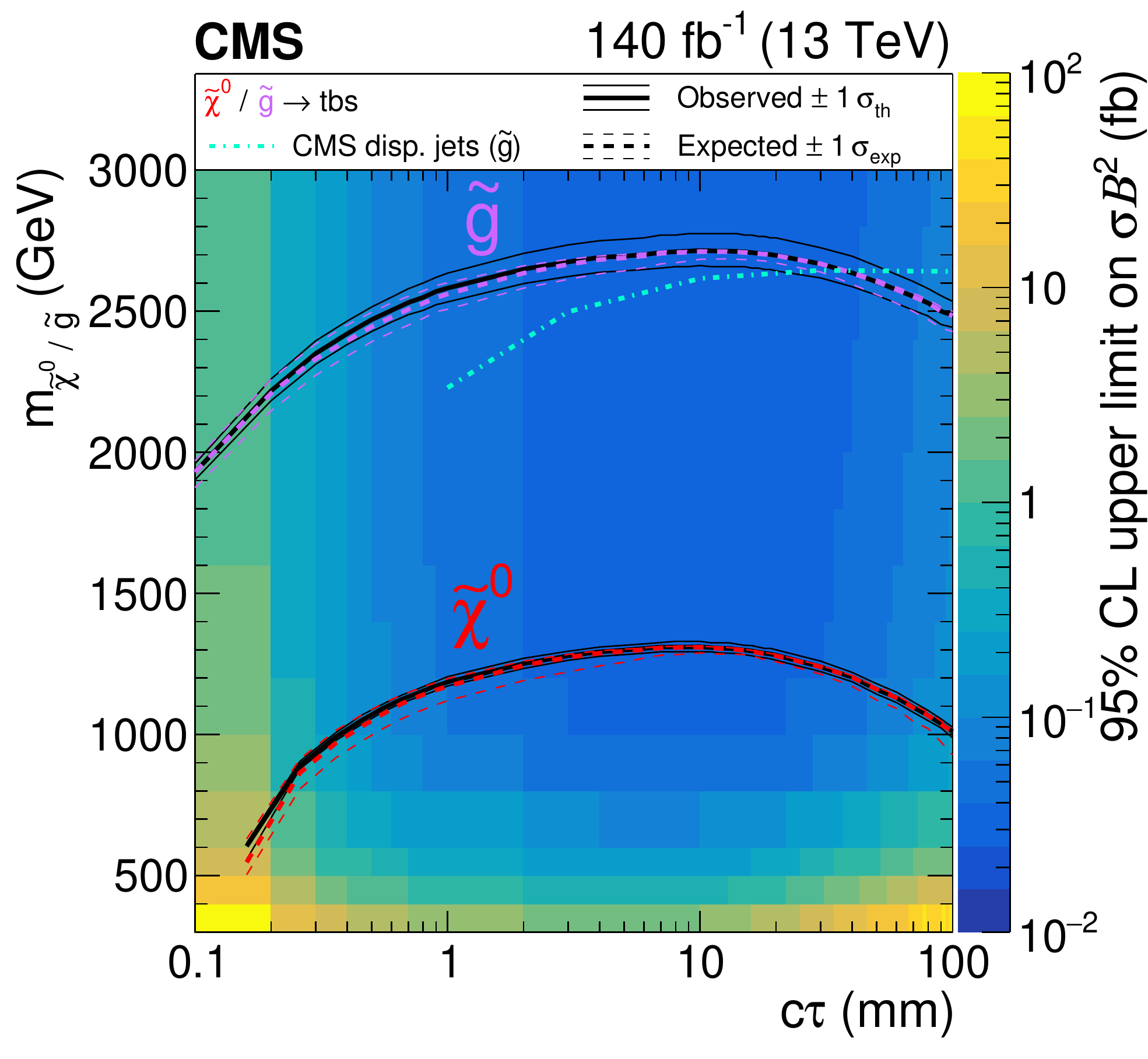}
\includegraphics[width=0.49\textwidth]{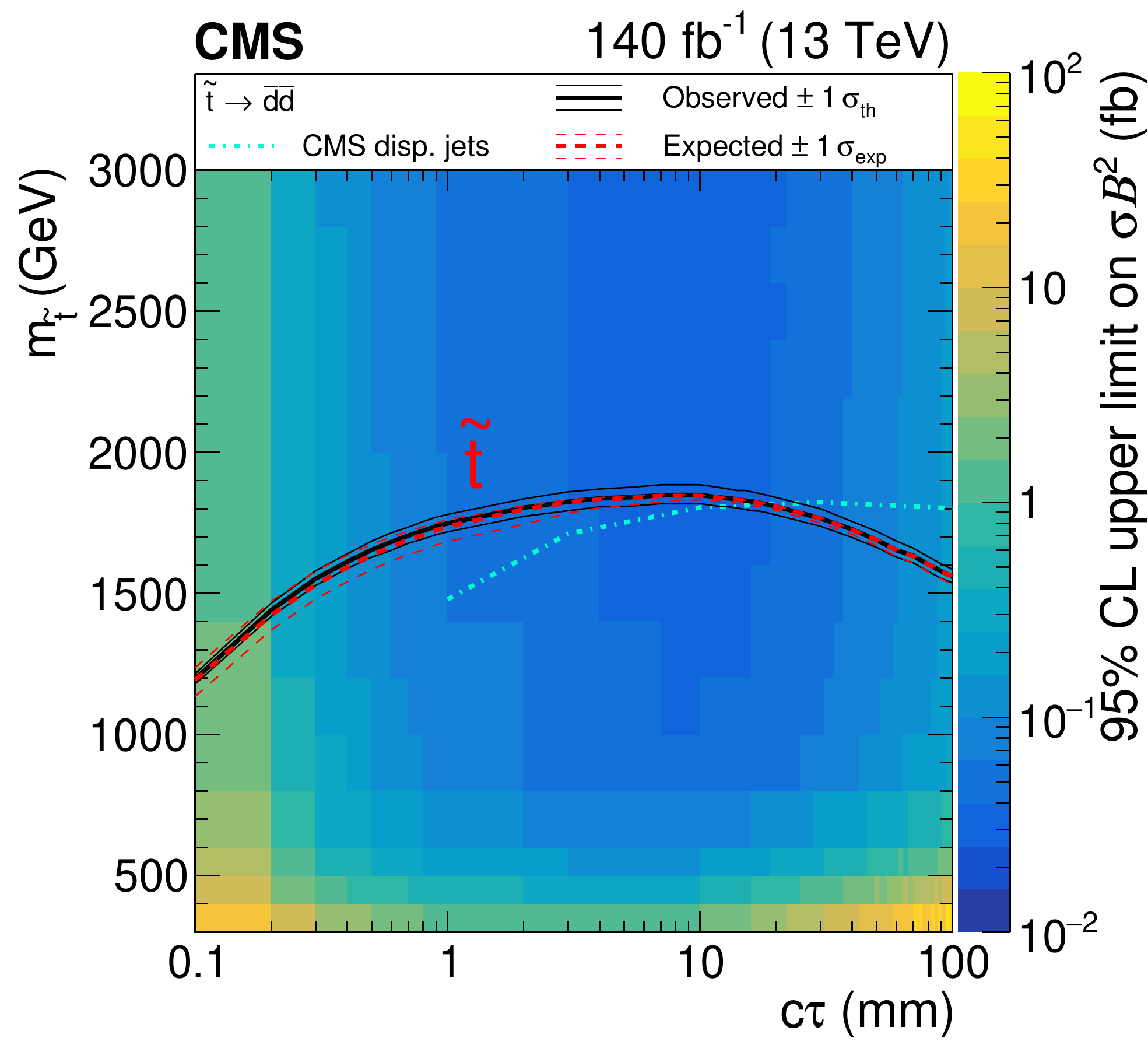}
\caption{Observed 95\% \CL upper limits on the product of cross section and branching fraction squared for the multijet (left) and dijet (right) signals, as a function of mass and $c\tau$. The overlaid mass-lifetime exclusion curves assume pair-production cross sections for the neutralino (red) and gluino (purple) in multijet signals and top squark cross sections for the dijet signals with 100\% branching fraction to each model's respective decay mode specified. The solid black (dashed colored) lines represent the observed (median expected) limits at 95\% \CL. The thin black lines represent the variation of the observed limit within theoretical uncertainties of the signal cross section. The thin dashed colored lines represent the region containing 68\% of the expected limit distribution under the background-only hypothesis. The observed limits from the CMS displaced jets search~\cite{Sirunyan:2020cao} are also shown in teal for comparison.}
\label{fig:limits}
\end{figure*}

The final fit combines the observed yields and background templates from the 2015--2018 data sets to achieve the full Run-2 result. The correlations between the old and new data sets are treated in the same way as the correlations between the 2017 and 2018 data sets. Figure~\ref{fig:limits} shows, for the full Run-2 result, the 95\% confidence level (\CL) upper limits on the product of the pair-production cross section and the square of the branching fraction ($\sigmaBsq$), as a function of mass and mean proper decay length. The exclusion curves overlaid use the neutralino production cross sections computed at next-to-leading-order (NLO) and next-to-leading-logarithm (NLL) precision in a limit of mass-degenerate higgsino states \PSGcpmDo, \PSGczDo, and \PSGczDt, with all of the other sparticles assumed to be heavy and decoupled~\cite{Fuks:2012qx,Fuks:2013vua}.  For the gluino and top squark, the mass-dependent production cross sections are computed at next-to-NLO$_{\mathrm{approx}}$ and next-to-NLL precision~\cite{gluinoxsec,Beenakker:2016gmf,Beneke:2016kvz}. For all models, we assume a 100\% branching fraction to the specified decay mode.

For the long-lived gluino, neutralino, and top squark in the RPV models described, pair-production cross sections larger than 0.08\fb are excluded for masses between 800 and 3000\GeV and mean proper decay lengths between 1 and 25\mm. For mean proper decay lengths between 0.6 and 90\mm, the data exclude gluino masses up to 2500\GeV; for mean proper decay lengths between 0.6 and 70\mm, the data exclude neutralino masses up to 1100\GeV; and for mean proper decay lengths between 0.4 and 80\mm, the data exclude top squark masses up to 1600\GeV. These exclusions are 250--300\GeV higher than in the previous analysis~\cite{CMS-EXO-17-018} and are the most stringent bounds on these models for mean proper decay lengths between 0.1 and 15\mm for all masses considered. In contrast, prompt searches have only excluded pair-produced prompt gluinos decaying into trijet final states for masses up to 1500\GeV and prompt top squarks decaying into dijet final states for masses up to 520\GeV~\cite{Sirunyan:2018rlj,Sirunyan:2018duw}. The potential long lifetime of the particle provides a handle to reduce the background and allows this search to have better sensitivity to larger masses.

\begin{figure*}[hbtp]
\centering
\includegraphics[width=0.45\textwidth]{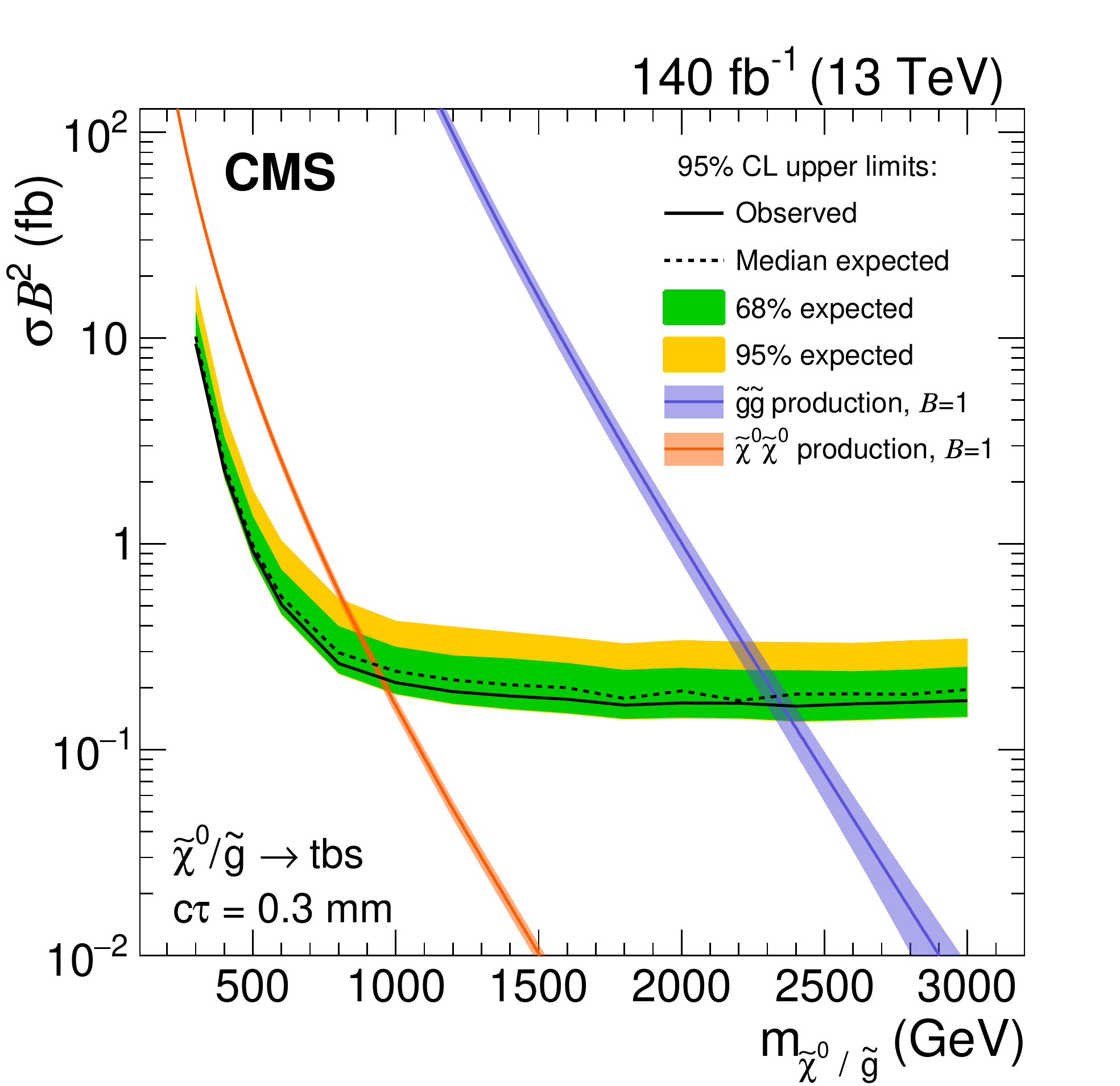}
\includegraphics[width=0.45\textwidth]{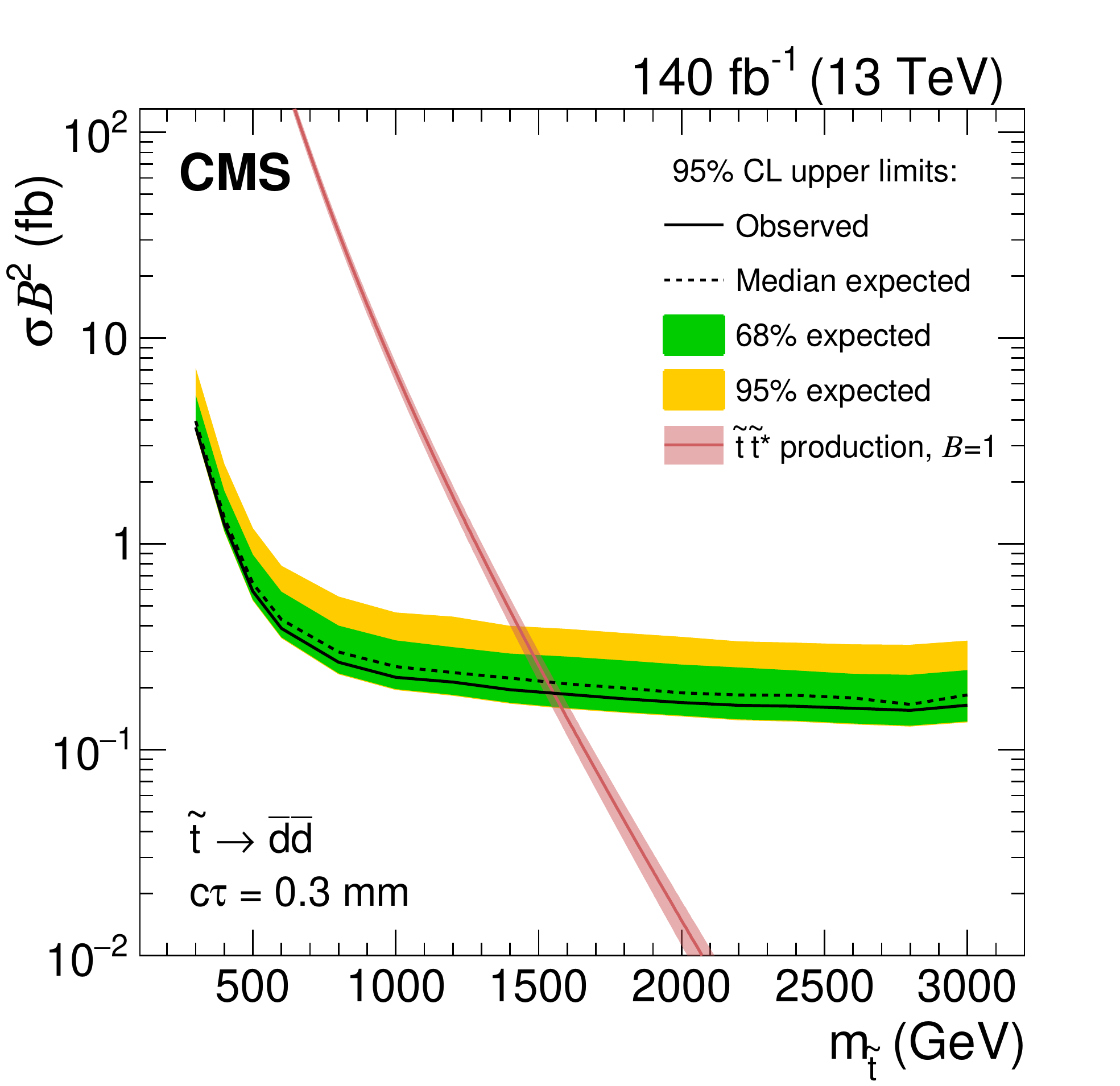}
\includegraphics[width=0.45\textwidth]{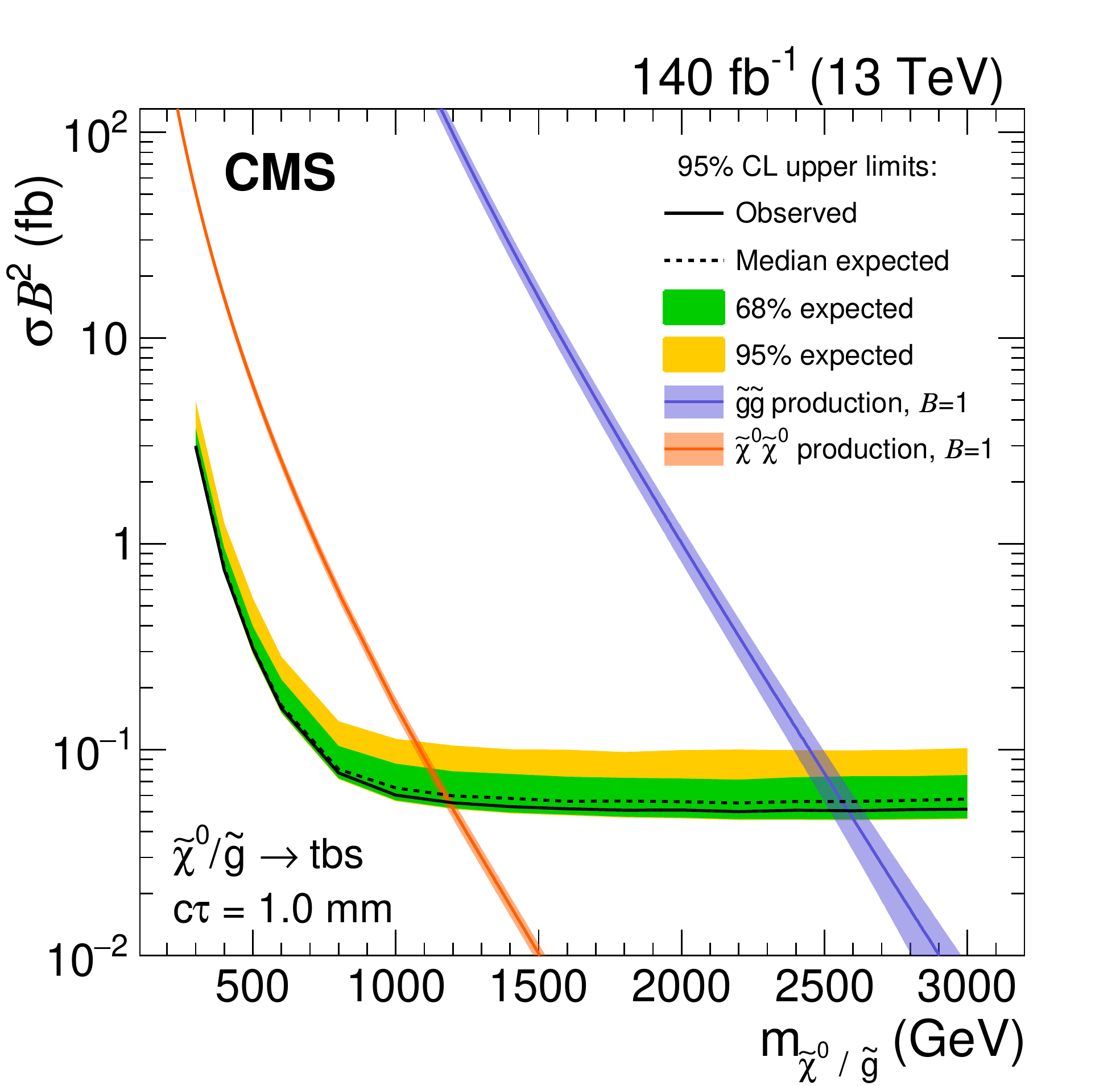}
\includegraphics[width=0.45\textwidth]{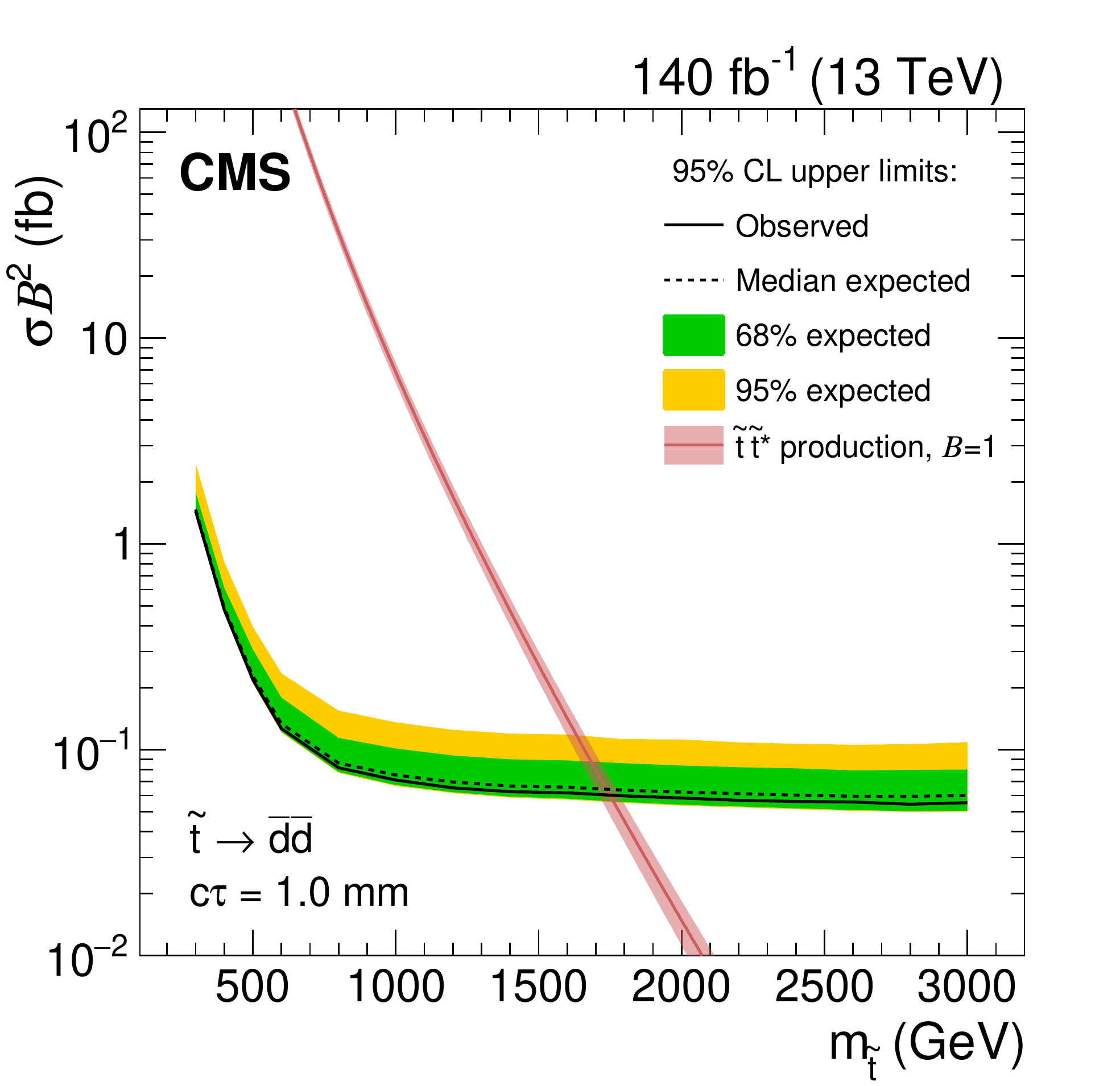}
\includegraphics[width=0.45\textwidth]{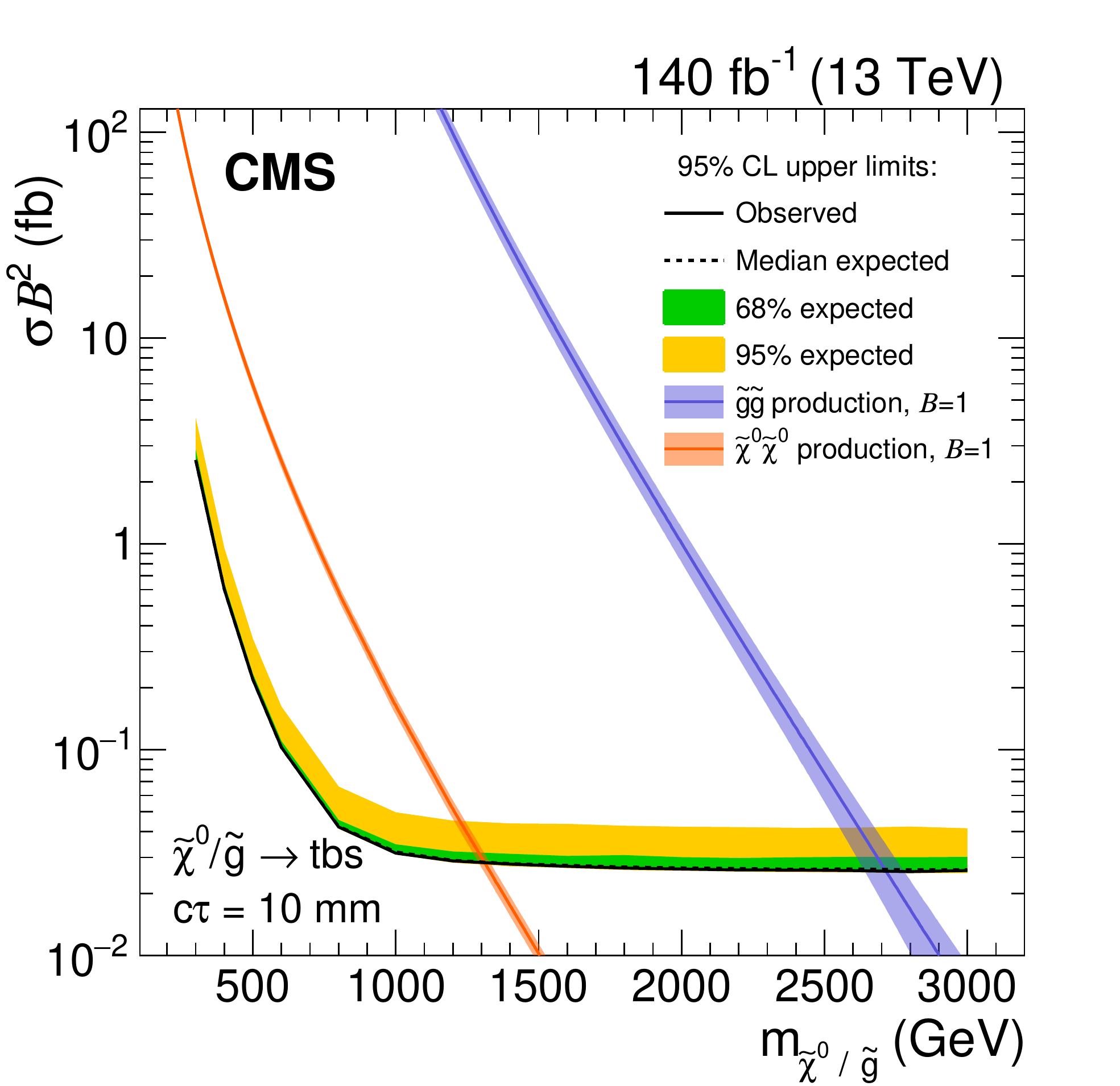}
\includegraphics[width=0.45\textwidth]{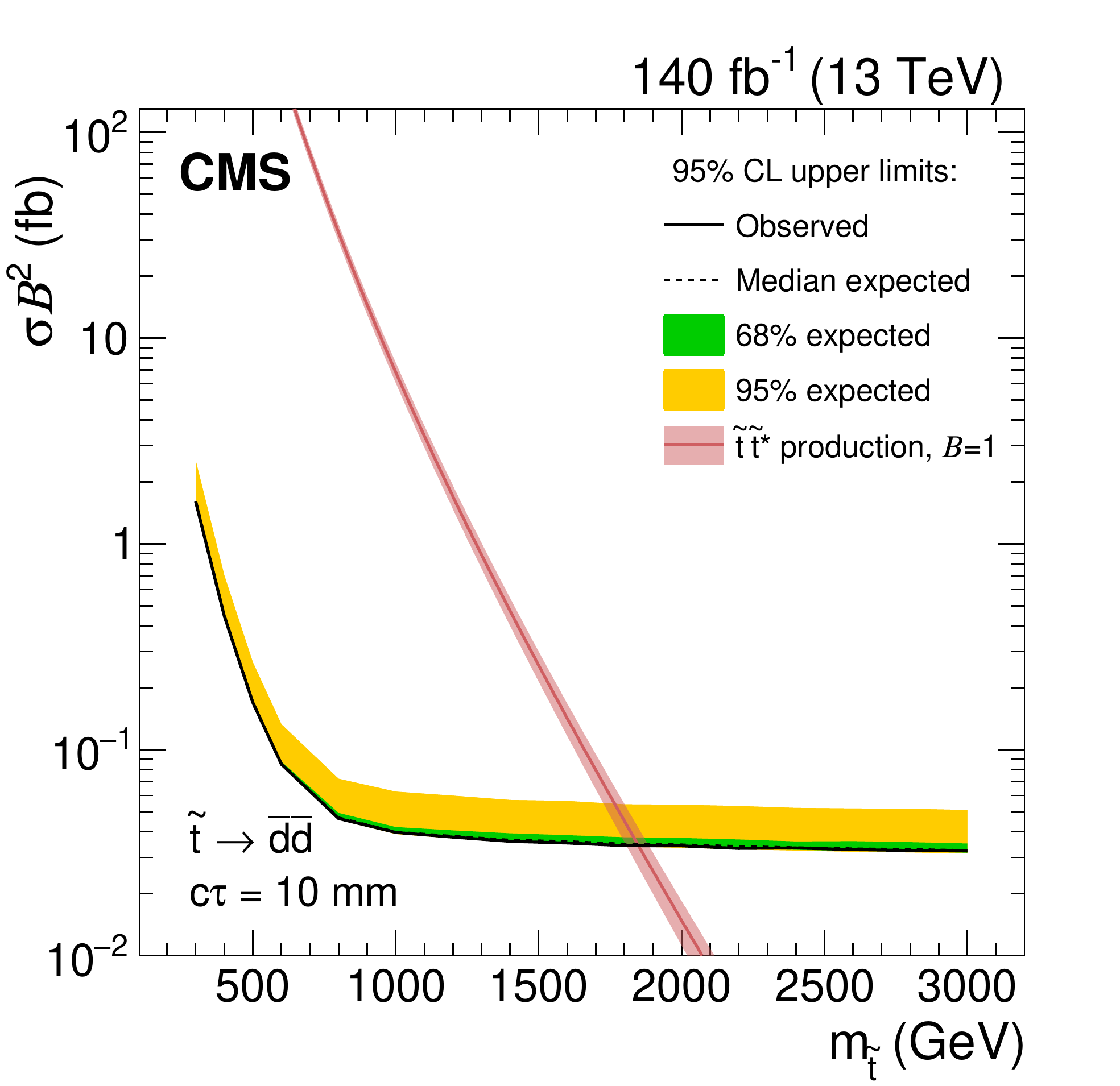}
\caption{Observed and expected 95\% \CL upper limits on the product of cross section 
and branching fraction squared, as a function of mass for multijet signals (left) and 
dijet signals (right), for a fixed $c\tau$ of 
0.3\mm (upper), 1\mm (middle), and 10\mm (lower) in the full Run-2 data set.  The neutralino and gluino pair 
production cross sections are shown for the multijet signals, and 
the top squark pair-production cross section is shown for the 
dijet signals.}
\label{fig:limits1d_vs_mass_run2}
\end{figure*}

\begin{figure*}[hbtp]
\centering
\includegraphics[width=0.45\textwidth]{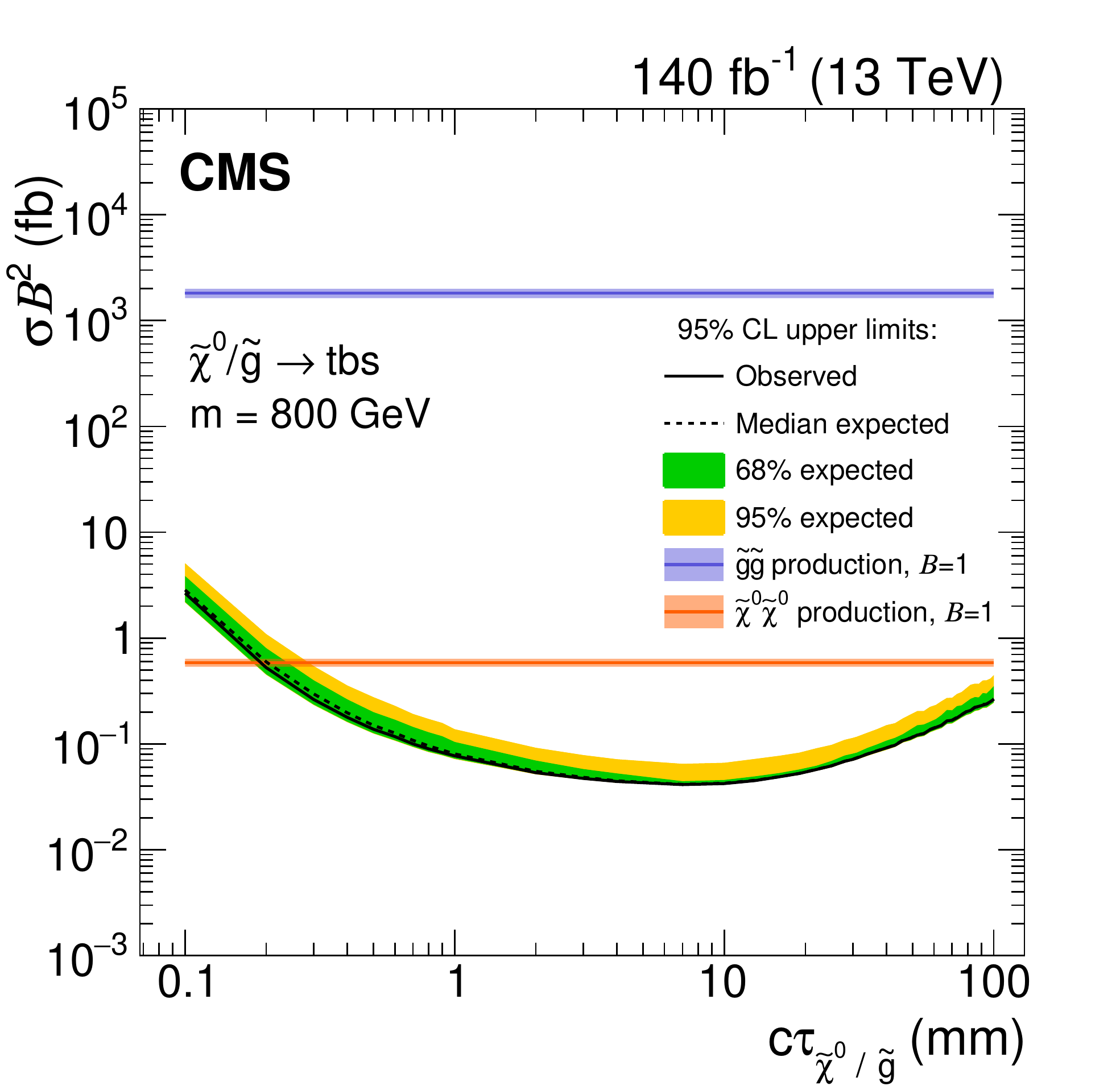}
\includegraphics[width=0.45\textwidth]{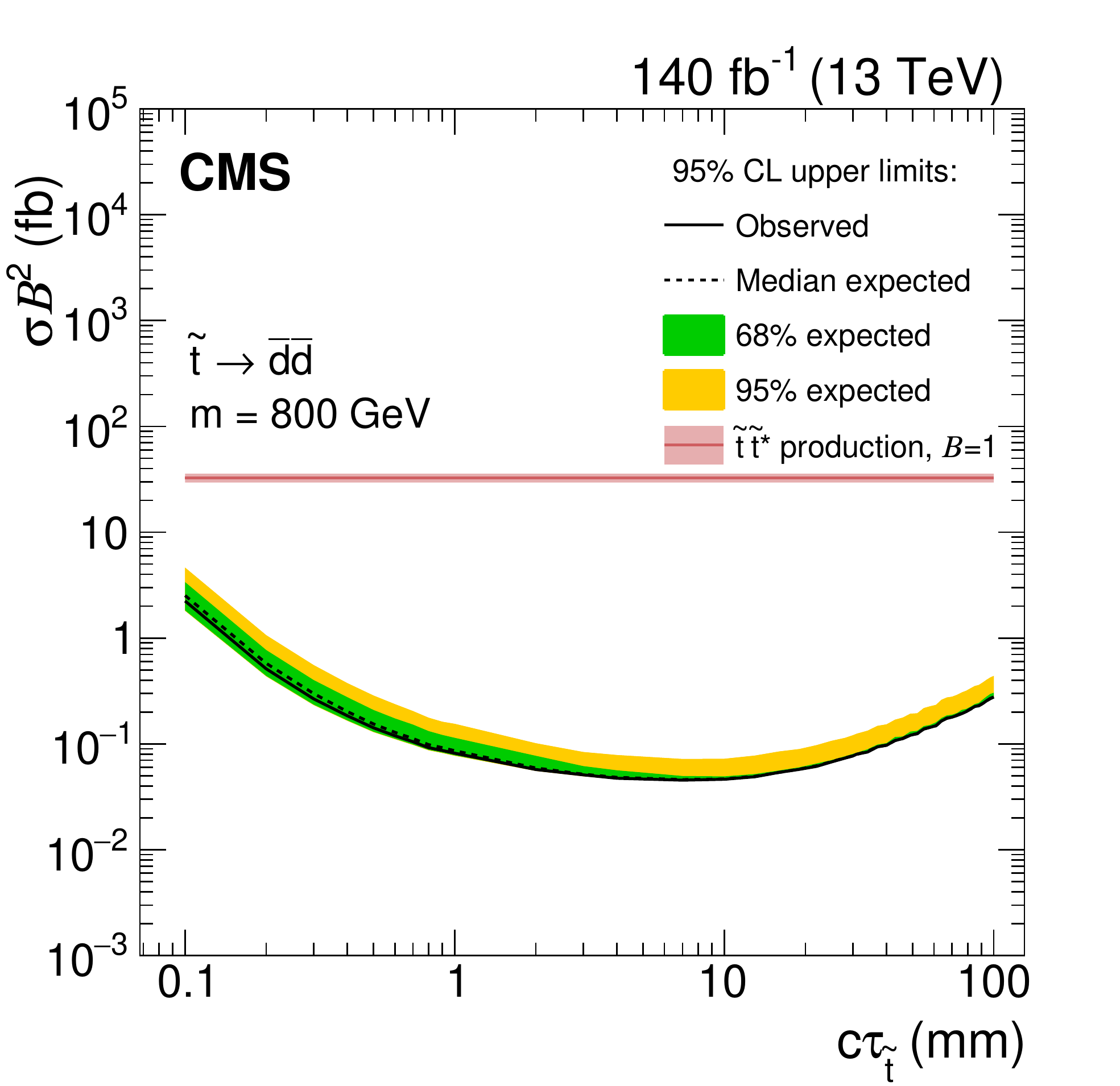}
\includegraphics[width=0.45\textwidth]{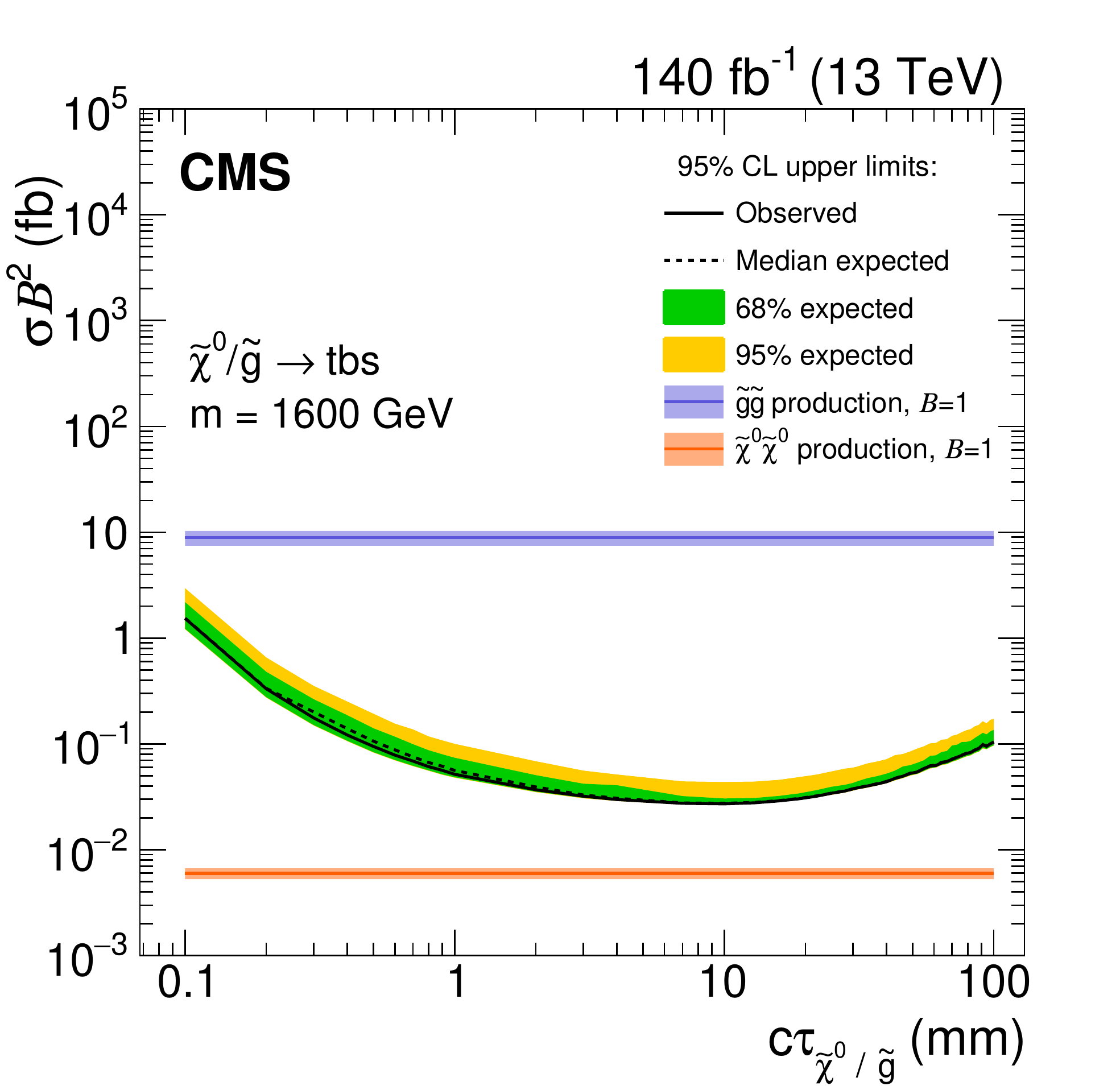}
\includegraphics[width=0.45\textwidth]{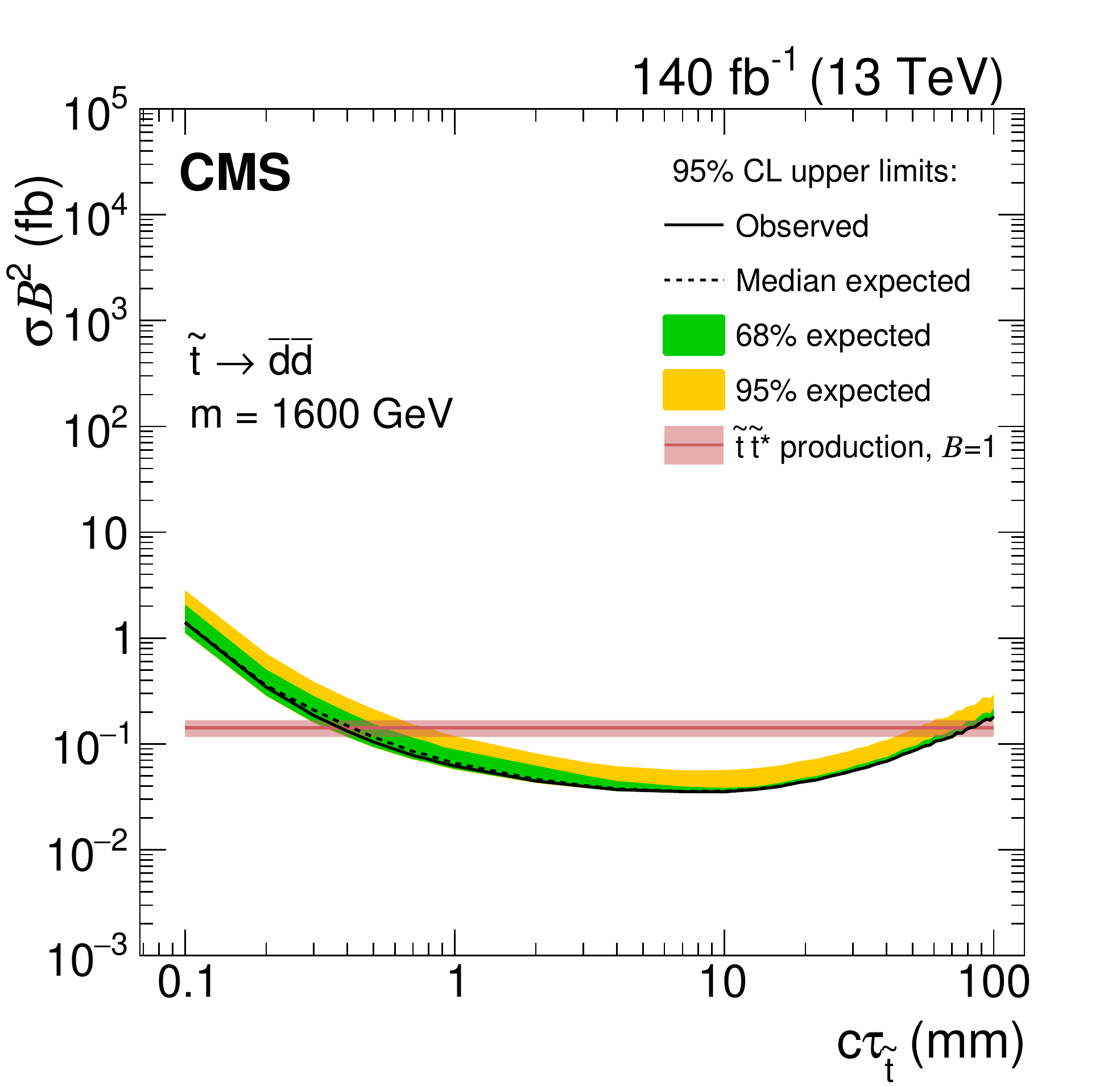}
\includegraphics[width=0.45\textwidth]{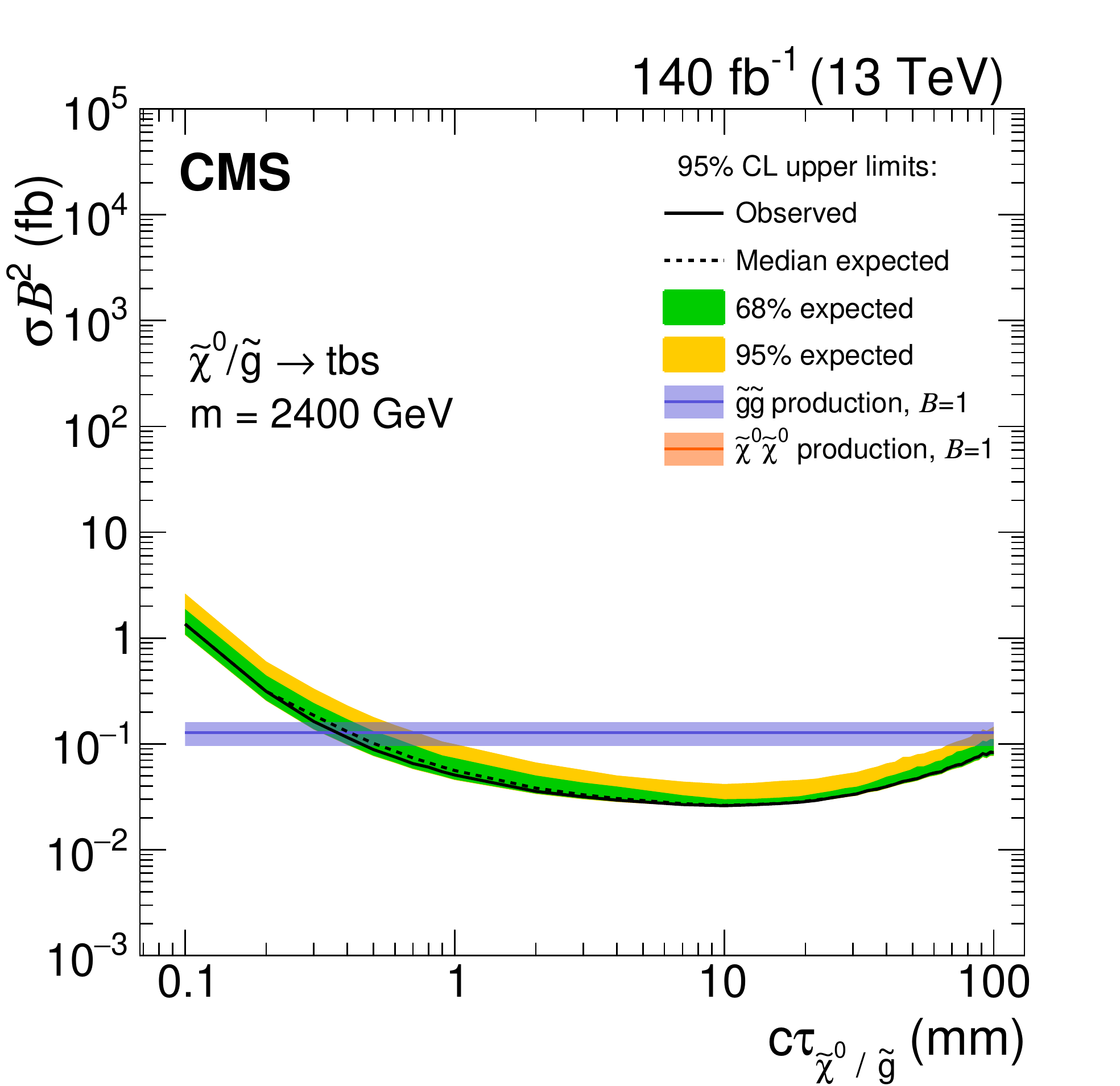}
\includegraphics[width=0.45\textwidth]{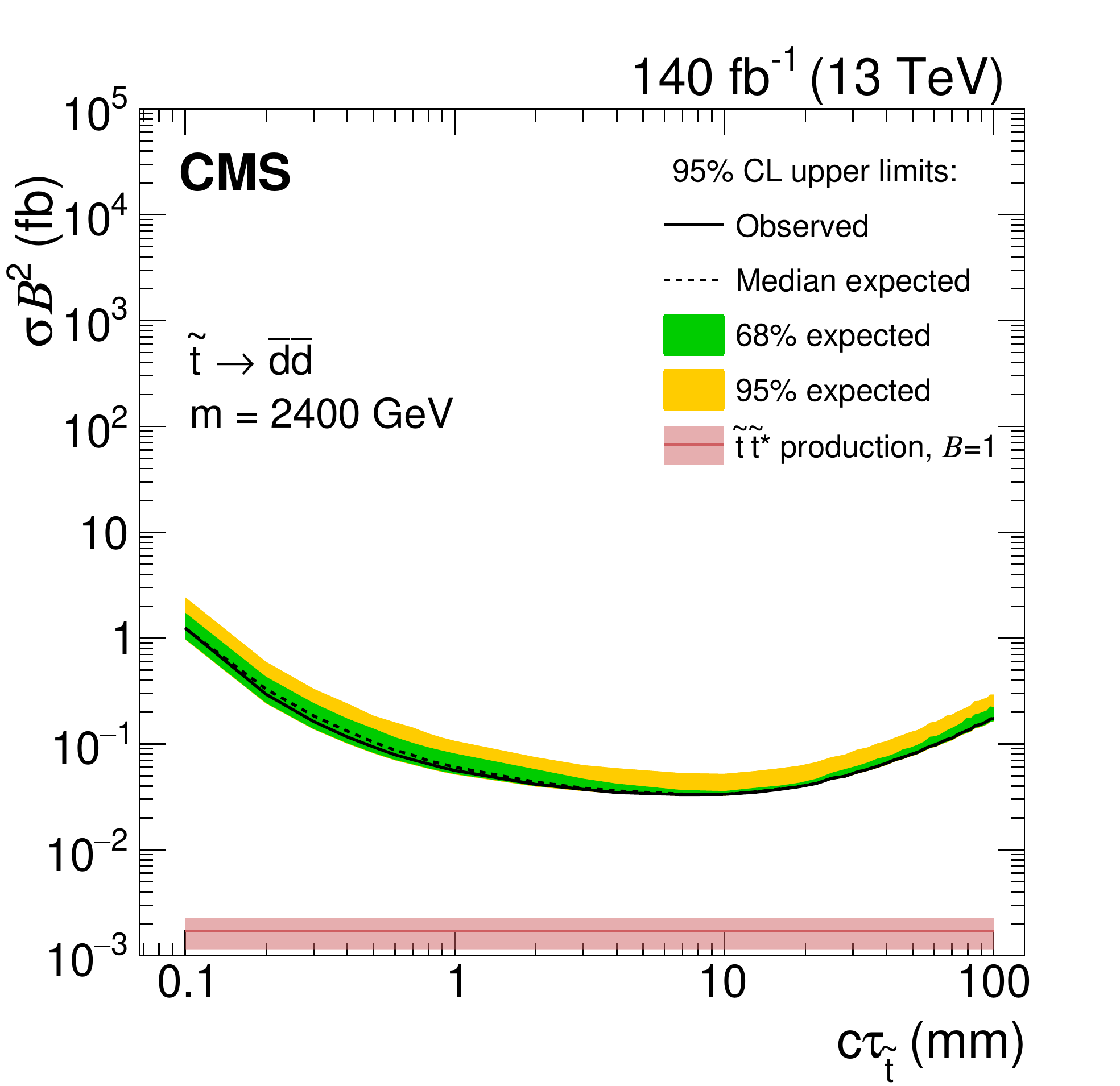}
\caption{Observed and expected 95\% \CL upper limits on the product of cross section 
and branching fraction squared, as a function of $c\tau$ for multijet signals (left) and 
dijet signals (right), for a fixed mass of 
800\GeV (upper), 1600\GeV (middle), and 2400\GeV (lower) in the full Run-2 data set. The neutralino and gluino pair 
production cross sections are shown for the multijet signals, and 
the top squark pair-production cross section is shown for the 
dijet signals. For $m=2400\GeV$, the expected neutralino cross section is $\approx 8\times 10^{-5}\fb$ and is not shown. }
\label{fig:limits1d_vs_ctau_run2}
\end{figure*}

Figure~\ref{fig:limits1d_vs_mass_run2} shows one-dimensional slices of the upper limit as a function of mass for several values of $c\tau$. Similarly, Fig.~\ref{fig:limits1d_vs_ctau_run2} shows the upper limit as a function of $c\tau$ for a selection of masses. 

At a specific signal point, with a gluino with mass 1600\GeV and mean proper decay length $c\tau$ of 10\mm, the computed 95\% \CL upper limit on $\sigmaBsq$ in the 2017 and 2018 data set alone is 0.04\fb, compared with the limit from the 2015 and 2016 data set of about 0.15\fb. The improvements arise primarily from the increase in statistical precision because of the increased integrated luminosity of \intlumiSecond compared with \intlumiFirst, in addition to the larger background estimated in the 2015 and 2016 data set due to the $\geq$5-track two-vertex event observed in that data set. By combining these two data sets, the 95\% \CL upper limit for the same signal point is lowered to 0.03\fb. 

These RPV SUSY models provide an illustrative interpretation of the data. However, the results may be applied to other models in which pairs of long-lived particles each decay into two or more jets in their final state. A set of instructions is contained in Appendix~\ref{app:reinterpretation}, providing a method for applying the results of this analysis to other signal models.

\section{Summary}
\label{sec:summary}
A search for pair-produced long-lived particles decaying into multijet and dijet final states in proton-proton collisions collected with the CMS detector at a center-of-mass energy of 13\TeV has been described. No events in the signal region in the 2017 and 2018 data sets, and no excess yield beyond the standard model prediction in the full Run-2 data set, which corresponds to an integrated luminosity of \intlumiTotal, are observed.  This analysis extends a previous CMS search that used the 2015 and 2016 data sets, with improvements in background rejection, background estimation techniques, and uncertainty estimation. 

At 95\% confidence level, upper limits are set on an $R$-parity violating (RPV) supersymmetry (SUSY) model in which a long-lived neutralino or gluino decays into a multijet final state with top, bottom, and strange quarks. Signal pair-production cross sections larger than 0.08\fb are excluded for long-lived neutralino, gluino, and top squark masses between 800 and 3000\GeV and mean proper decay lengths between 1 and 25\mm. For the range of mean proper decay lengths between 0.6 and 90\mm, the data exclude gluino masses up to 2500\GeV. For the case where the lightest SUSY particle is a neutralino, the data exclude neutralino masses up to 1100\GeV for mean proper decay lengths between 0.6 and 70\mm. Additionally, limits are placed for an RPV SUSY model in which a long-lived top squark decays into a dijet final state with two down-type quarks. The data exclude top squark masses up to 1600\GeV for mean proper decay lengths between 0.4 and 80\mm.  These results, which supersede those in Ref.~\cite{CMS-EXO-17-018}, are the most stringent bounds on these models for mean proper decay lengths between 0.1 and 15\mm for all masses considered, and
complement the results of the CMS displaced jets search~\cite{Sirunyan:2020cao}. While the search directly constrains these two RPV SUSY models, the techniques and methodology are generic and, as described in the Appendix, the results are applicable to other models of pair-produced long-lived particles that decay into jets. 

\newpage

\begin{acknowledgments}
  We congratulate our colleagues in the CERN accelerator departments for the excellent performance of the LHC and thank the technical and administrative staffs at CERN and at other CMS institutes for their contributions to the success of the CMS effort. In addition, we gratefully acknowledge the computing centers and personnel of the Worldwide LHC Computing Grid and other centers for delivering so effectively the computing infrastructure essential to our analyses. Finally, we acknowledge the enduring support for the construction and operation of the LHC, the CMS detector, and the supporting computing infrastructure provided by the following funding agencies: BMBWF and FWF (Austria); FNRS and FWO (Belgium); CNPq, CAPES, FAPERJ, FAPERGS, and FAPESP (Brazil); MES (Bulgaria); CERN; CAS, MoST, and NSFC (China); MINCIENCIAS (Colombia); MSES and CSF (Croatia); RIF (Cyprus); SENESCYT (Ecuador); MoER, ERC PUT and ERDF (Estonia); Academy of Finland, MEC, and HIP (Finland); CEA and CNRS/IN2P3 (France); BMBF, DFG, and HGF (Germany); GSRT (Greece); NKFIA (Hungary); DAE and DST (India); IPM (Iran); SFI (Ireland); INFN (Italy); MSIP and NRF (Republic of Korea); MES (Latvia); LAS (Lithuania); MOE and UM (Malaysia); BUAP, CINVESTAV, CONACYT, LNS, SEP, and UASLP-FAI (Mexico); MOS (Montenegro); MBIE (New Zealand); PAEC (Pakistan); MSHE and NSC (Poland); FCT (Portugal); JINR (Dubna); MON, RosAtom, RAS, RFBR, and NRC KI (Russia); MESTD (Serbia); SEIDI, CPAN, PCTI, and FEDER (Spain); MOSTR (Sri Lanka); Swiss Funding Agencies (Switzerland); MST (Taipei); ThEPCenter, IPST, STAR, and NSTDA (Thailand); TUBITAK and TAEK (Turkey); NASU (Ukraine); STFC (United Kingdom); DOE and NSF (USA).

  \hyphenation{Rachada-pisek} Individuals have received support from the Marie-Curie program and the European Research Council and Horizon 2020 Grant, contract Nos.\ 675440, 724704, 752730, 765710 and 824093 (European Union); the Leventis Foundation; the Alfred P.\ Sloan Foundation; the Alexander von Humboldt Foundation; the Belgian Federal Science Policy Office; the Fonds pour la Formation \`a la Recherche dans l'Industrie et dans l'Agriculture (FRIA-Belgium); the Agentschap voor Innovatie door Wetenschap en Technologie (IWT-Belgium); the F.R.S.-FNRS and FWO (Belgium) under the ``Excellence of Science -- EOS" -- be.h project n.\ 30820817; the Beijing Municipal Science \& Technology Commission, No. Z191100007219010; the Ministry of Education, Youth and Sports (MEYS) of the Czech Republic; the Deutsche Forschungsgemeinschaft (DFG), under Germany's Excellence Strategy -- EXC 2121 ``Quantum Universe" -- 390833306, and under project number 400140256 - GRK2497; the Lend\"ulet (``Momentum") Program and the J\'anos Bolyai Research Scholarship of the Hungarian Academy of Sciences, the New National Excellence Program \'UNKP, the NKFIA research grants 123842, 123959, 124845, 124850, 125105, 128713, 128786, and 129058 (Hungary); the Council of Science and Industrial Research, India; the Ministry of Science and Higher Education and the National Science Center, contracts Opus 2014/15/B/ST2/03998 and 2015/19/B/ST2/02861 (Poland); the National Priorities Research Program by Qatar National Research Fund; the Ministry of Science and Higher Education, project no. 0723-2020-0041 (Russia); the Programa Estatal de Fomento de la Investigaci{\'o}n Cient{\'i}fica y T{\'e}cnica de Excelencia Mar\'{\i}a de Maeztu, grant MDM-2015-0509 and the Programa Severo Ochoa del Principado de Asturias; the Thalis and Aristeia programs cofinanced by EU-ESF and the Greek NSRF; the Rachadapisek Sompot Fund for Postdoctoral Fellowship, Chulalongkorn University and the Chulalongkorn Academic into Its 2nd Century Project Advancement Project (Thailand); the Kavli Foundation; the Nvidia Corporation; the SuperMicro Corporation; the Welch Foundation, contract C-1845; and the Weston Havens Foundation (USA).
\end{acknowledgments}

\bibliography{auto_generated}

\newpage

\numberwithin{table}{section}
\appendix
\section{Applying the results to different models}
\label{app:reinterpretation}
While the search presented specifically addresses two models of RPV SUSY, the results may be applied to other models in which the pair-produced long-lived particles each decay into two or more jets in the final state. Conversion of the upper limit on signal events to an upper limit on the corresponding signal cross section depends on the reconstruction efficiency for that model. In this section, we present a set of generator-level selection requirements that, when applied, approximate the reconstruction-level efficiency of this analysis and allow for reinterpretation of the results. 

Event selection is based on the properties of generated jets in the event, as well as quarks and leptons produced in the decays of the long-lived particles. We assume that the generated jets are clustered from all final-state particles, excluding neutrinos, using the anti-\kt algorithm with a distance parameter of 0.4. A jet is rejected if the fraction of energy shared by electrons is greater than 0.9, or similarly if the muon energy fraction is greater than 0.8. We apply additional kinematic requirements at the parton level to the \PQu, \PQd, \PQs, {\cPqc}, and \PQb quarks in addition to the electron, muon, and tau leptons from the long-lived particle decay. These daughter particles must have a transverse impact parameter with respect to the origin of at least 0.1\mm. To be selected, generated jets and the daughter particles must satisfy $\pt > 20\GeV$ and $\abs{\eta}<2.5$.

The following generator-level selection requirements approximate the reconstruction-level criteria: 
\begin{itemize}
\item Each event must contain at least four generated jets.
\item $\HT$ must be greater than 1200\GeV, where \HT is the scalar \pt sum of generated jets with $\pt > 40\GeV$.
\item The distance of the decay point from the origin in the $x$-$y$ plane of each generated long-lived particle must be within 0.1 and 20\mm.
\item The $\Sigma\pt$ of the daughter particles of each long-lived particle must exceed 350\GeV to ensure sufficiently small uncertainty in \dbv and sufficiently large number of tracks per vertex. However, if the daughter particle is a bottom quark, its $\Sigma\pt$ is scaled down by a factor of 0.65. This corrects for reduced reconstruction efficiency for bottom quarks due to their lifetime, which can inhibit the association of their decay products with the reconstructed vertex.
\item The transverse distance between the decay points of each long-lived particle must be greater than 0.4\mm.
\end{itemize}

Following this prescription, the generator-level efficiency approximates the reconstruction-level efficiency with 20\% accuracy for a wide variety of models. Finally, to correct for differences between data and simulation, the event yields must be scaled by the data-to-simulation efficiency correction factors provided in Table~\ref{tab:eventSFs}, which approximate those described in Section~\ref{sec:sigeff}. This was tested for models with both dijet and multijet final states for masses of 400--3000\GeV and mean proper decay lengths of 0.1--30\mm. This prescription has been validated only for models with efficiency greater than 10\%. 

\begin{table*}[htbp!]
\centering
\topcaption{Data-to-simulation efficiency correction factors, shown for multijet and dijet signal topologies in several ranges of $c\tau$. Note that these correction factors account for the two long-lived particles in the simulated events, and are therefore the total correction factors used to scale event yields rather than the correction factors one would apply to individual vertices.}
\begin{scotch}{lcccc}
& 0.1--0.3\mm & 0.3--1\mm & 1--10\mm & 10--100\mm \\
\hline
Correction factor for multijet signals & 0.92 & 0.94 & 0.97 & 0.98 \\
Correction factor for dijet signals    & 0.75 & 0.79 & 0.82 & 0.84 \\
\end{scotch}
\label{tab:eventSFs}
\end{table*}
\cleardoublepage \section{The CMS Collaboration \label{app:collab}}\begin{sloppypar}\hyphenpenalty=5000\widowpenalty=500\clubpenalty=5000\vskip\cmsinstskip
\textbf{Yerevan Physics Institute, Yerevan, Armenia}\\*[0pt]
A.M.~Sirunyan$^{\textrm{\dag}}$, A.~Tumasyan
\vskip\cmsinstskip
\textbf{Institut f\"{u}r Hochenergiephysik, Wien, Austria}\\*[0pt]
W.~Adam, T.~Bergauer, M.~Dragicevic, A.~Escalante~Del~Valle, R.~Fr\"{u}hwirth\cmsAuthorMark{1}, M.~Jeitler\cmsAuthorMark{1}, N.~Krammer, L.~Lechner, D.~Liko, I.~Mikulec, F.M.~Pitters, J.~Schieck\cmsAuthorMark{1}, R.~Sch\"{o}fbeck, M.~Spanring, S.~Templ, W.~Waltenberger, C.-E.~Wulz\cmsAuthorMark{1}, M.~Zarucki
% ...
\vskip\cmsinstskip
\dag: Deceased\\
1:  Also at TU Wien, Wien, Austria\\
% ...
\end{sloppypar}
%%% END EDITABLE REGION %%%
% skeleton_end
\end{document}